\documentclass{emulateapj}

\usepackage{apjfonts}
\usepackage{natbib}
\usepackage{graphics}
\usepackage{multirow}
\usepackage{amsbsy}
\usepackage{mathrsfs}
\usepackage{color}
\usepackage{url}

\def\bicep{{\sc Bicep1}}
\def\bolocam{{\sc Bolocam}}

\def\biceptwo{{\sc Bicep2}}
\def\bicepthree{{\sc Bicep3}}

\def\planck{{\it Planck}}
\def\quiet{{ QUIET}}

\def\acbar{{\sc Acbar}}
\def\QUAD{{\sc QUaD}}

\def\boom{{\sc Boomerang}}
\def\wmap{{\sc Wmap}}
\def\spt{{\sc SPT}}

\def\keckarray{Keck Array}

\def\camb{{\tt CAMB}}

\def\synfast{{\tt synfast}}
\def\healpix{{\tt HEALPix}}

\def\spice{{\tt Spice}}

\def\muK{~\mu{\rm K}}
\def\mukcmb{\mu\mathrm{K}_{\mathrm{\mbox{\small\sc cmb}}}}

\def\deg{^\circ}
\def\emode{$E$-mode}
\def\bmode{$B$-mode}
\def\BB{$BB$}

\def\lcdm{$\Lambda$CDM}

\DeclareMathAlphabet\mathbfcal{OMS}{cmsy}{b}{n} 

\defcitealias{chiang_2010}{C10}
\defcitealias{takahashi_2010}{T10}
\defcitealias{pryke_2009}{P09}

\def\aap{Astr. \& Astroph.\ }

\def\apjl{Astrophys.\ J.\ Lett.\ }
\def\apjs{Astrophys.\ J.\ Suppl.\ }

\shorttitle{CMB POLARIZATION FROM THREE YEARS OF BICEP1}
\shortauthors{BARKATS ET AL.}
\submitted{ApJ 783, 67, (2014)}

\begin{document}

\title{Degree-Scale CMB Polarization Measurements from Three Years of BICEP1 Data}

\author{\bicep\ Collaboration---D.~Barkats\altaffilmark{1}}
\author{R.~Aikin\altaffilmark{2}}
\author{C.~Bischoff\altaffilmark{3}}
\author{I.~Buder\altaffilmark{3}}
\author{J.~P.~Kaufman\altaffilmark{8}}
\author{B.~G.~Keating\altaffilmark{8}}
\author{J.~M.~Kovac\altaffilmark{3}}
\author{M.~Su\altaffilmark{4,5}}
\author{P.~A.~R.~Ade\altaffilmark{6}}
\author{J.~O.~Battle\altaffilmark{7}}
\author{E.~M.~Bierman\altaffilmark{8}}
\author{J.~J.~Bock\altaffilmark{2,7}}
\author{H.~C.~Chiang\altaffilmark{9}}
\author{C.~D.~Dowell\altaffilmark{7}}
\author{L.~Duband\altaffilmark{10}}
\author{J.~Filippini\altaffilmark{2}}
\author{E.~F.~Hivon\altaffilmark{11}}
\author{W.~L.~Holzapfel\altaffilmark{12}}
\author{V.~V.~Hristov\altaffilmark{2}}
\author{W.~C.~Jones\altaffilmark{13}}
\author{C.~L.~Kuo\altaffilmark{14,15}}
\author{E.~M.~Leitch\altaffilmark{17}}
\author{P.~V.~Mason\altaffilmark{2}}
\author{T.~Matsumura\altaffilmark{19}}
\author{H.~T.~Nguyen\altaffilmark{7}}
\author{N.~Ponthieu\altaffilmark{16}}
\author{C.~Pryke\altaffilmark{18}}
\author{S.~Richter\altaffilmark{2}}
\author{G.~Rocha\altaffilmark{2,7}}
\author{C.~Sheehy\altaffilmark{17}}
\author{S.~S.~Kernasovskiy\altaffilmark{14,15}}
\author{Y.~D.~Takahashi\altaffilmark{12}}
\author{J.~E.~Tolan\altaffilmark{14,15}}
\author{K.~W.~Yoon\altaffilmark{14,15}}

\altaffiltext{1}{Joint ALMA Observatory, ESO, Santiago, Chile, {\bf dbarkats@alma.cl}}
\altaffiltext{2}{Department of Physics, California Institute of Technology, Pasadena, CA 91125, USA}
\altaffiltext{3}{Harvard-Smithsonian Center for Astrophysics, 60 Garden Street MS 42, Cambridge, MA 02138, USA}
\altaffiltext{4}{Department of Physics, Massachusetts Institute of Technology, 77 Massachusetts Avenue, Cambridge, MA, USA}
\altaffiltext{5}{MIT-Kavli Center for Astrophysics and Space Research, 77 Massachusetts Avenue, Cambridge, MA, USA}
\altaffiltext{6}{Department of Physics and Astronomy, University of Wales, Cardiff, CF24 3YB, Wales, UK}
\altaffiltext{7}{Jet Propulsion Laboratory, Pasadena, CA 91109, USA}
\altaffiltext{8}{Department of Physics, University of California at San Diego, La Jolla, CA 92093, USA}
\altaffiltext{9}{Astrophysics and Cosmology Research Unit, University of KwaZulu-Natal, Durban, South Africa}
\altaffiltext{10}{SBT, Commissariat \`a l'Energie Atomique, Grenoble, France}
\altaffiltext{11}{Institut d'Astrophysique de Paris, Paris, France}
\altaffiltext{12}{Department of Physics, University of California at Berkeley, Berkeley, CA 94720, USA}
\altaffiltext{13}{Department of Physics, Princeton University, Princeton, NJ, 08544, USA}
\altaffiltext{14}{Stanford University, Palo Alto, CA 94305, USA}
\altaffiltext{15}{Kavli Institute for Particle Astrophysics and Cosmology (KIPAC), Sand Hill Road 2575, Menlo Park, CA 94025, USA}
\altaffiltext{16}{Institut d'Astrophysique Spatiale, Universit\'e Paris-Sud, Orsay, France}
\altaffiltext{17}{University of Chicago, Chicago, IL 60637, USA}
\altaffiltext{18}{Department of Physics, University of Minnesota, Minneapolis, MN, 55455, USA}
\altaffiltext{19}{High Energy Accelerator Research Organization (KEK), Ibaraki, 305-0801, Japan}

\begin{abstract}

\noindent
BICEP1 is a millimeter-wavelength telescope designed specifically to measure the inflationary \bmode\ polarization of the Cosmic Microwave Background (CMB) at degree angular scales. We present results from an analysis of the data acquired during three seasons of observations at the South Pole (2006 to 2008). This work extends the two-year result published in \citet{chiang_2010}, with 
additional data from the third season and relaxed detector-selection criteria. This 
analysis also introduces a more comprehensive estimation of band-power window functions, improved likelihood estimation methods and a new technique for deprojecting monopole temperature-to-polarization leakage which reduces this class of systematic uncertainty to a negligible level.  We 
present maps of temperature, $E$- and $B$-mode polarization, and their associated angular power spectra. The improvement in the map noise level and polarization 
spectra error bars are consistent with the 52\% increase in integration time relative to~\citet{chiang_2010}.  We confirm both self-consistency of the polarization data and consistency with the 
two-year results. We measure   
the angular power spectra at $21 \leq \ell \leq 335 $ and find that the $EE$ spectrum is consistent with Lambda Cold Dark Matter (\lcdm) cosmology, with the first acoustic peak of the $EE$ spectrum now detected at 
15$\sigma$.
The $BB$ spectrum remains consistent with zero. From {\bmode}s only, we constrain the tensor-to-scalar ratio to $r = 0.03^{+0.27}_{-0.23}$, or $r < 0.70$ at 95\% confidence level. 
\end{abstract}


\section{Introduction}
 
In recent years, observational cosmology has produced enormous advances
in our understanding of the Universe. Observations of the
Cosmic Microwave Background (CMB) have played a central role in establishing what is now known 
as the standard cosmological model. Measurements of the CMB
temperature anisotropies have reached sub-percent 
precision over the whole sky at a range of angular scales down to few-arcminute resolution. 
The angular power spectrum of temperature anisotropies has yielded
tight constraints on the basic parameters of that
cosmological model, referred to as Lambda Cold Dark Matter (\lcdm). 
Although temperature anisotropy experiments 
\citep[e.g.]{story_2012, hinshaw_2013, planck_2013_xvi, sievers_2013} 
continue to test the validity of \lcdm, 
the model by itself 
offers no solution to the following mysteries: the 
high degree of flatness of the universe, the
apparent large-scale correlations that suggest a larger particle
horizon than allowed by the standard Big Bang scenario, the
nearly--scale-invariant spectrum of initial perturbations, and the lack of relic magnetic  monopoles. 

The inflationary scenario  was
proposed as an explanation to these observed properties of the
universe~\citep[for review, see][]{liddle_2000}.
One as-yet-unobserved prediction of inflation is a stochastic gravitational
wave background that would imprint its signature on the
anisotropies of the CMB. 
The most powerful method to search for this signature is to constrain the curl-mode (\bmode) polarization pattern of the CMB at degree angular scales~\citep{seljak_96,seljak_1997,Kamionkowski_1997}.

The CMB is polarized at the 10\% level
due to Thomson scattering at the
surface of last scattering. The density perturbations that give rise
to the temperature anisotropies also cause the plasma to flow along
gradients of this density field and so can only create gradient-mode, or \emode, polarization with zero curl \citep{hu_1997}.
Since the first detection of \emode\ polarization in 2002 \citep{kovac_2002},
several other experiments have refined the characterization of the $EE$ and $TE$ spectra \citep{montroy_2006, sievers_2007, wu_2007,  bischoff_2008, brown_2009, chiang_2010, quiet_2011, quiet_2012, bennett_2013}.
So far all have unambiguously confirmed the basic tenets of \lcdm. 

In addition to being a prediction of inflation, {\bmode}s can also be 
generated through gravitational lensing of {\emode}s, producing a signature that
is observationally distinct from inflationary {\bmode}s and
peaks at smaller angular scales.
The lensing {\bmode} polarization has recently been detected, using cross-correlations formed
with an external lensing template \citep{hanson_2013}. 
The inflationary {\bmode} pattern in the CMB polarization still remains elusive. 
A detection of primordial {\bmode} polarization would provide strong support to the inflationary scenario.

The amplitude of the {\bmode} signal is
parametrized by the tensor-to-scalar ratio $r$. 
The best constraints on $r$ are currently derived from CMB temperature
anisotropies: $r < 0.11$ at 95\% confidence for models that add only tensors to \lcdm,
or $r < 0.23$ for models allowing running of the spectral index \citep{story_2012,planck_2013_xxii}.
Cosmic variance limits further
improvements on $r$ constraints using temperature anisotropies alone.

The best limit on $r$ using only \bmode\ polarization is $r < 0.72$ at 95\% CL, set using \bicep\ data \citep[hereafter \citetalias{chiang_2010}]{chiang_2010}. This previous result only used the
first two years of observations and conservative detector-selection criteria.  In
this paper, we report measurements of the CMB polarization power
spectra and improved constraints on $r$, using data from all three years, as well as relaxed detector-selection criteria. We also
present an updated data-analysis method which includes an
improved noise model, a more sophisticated calculation of the 
band-power window functions, and new likelihood estimation techniques. 
Another unique addition to this analysis is a
deprojection filter which suppresses instrumental temperature-to-polarization   
leakage from relative gain miscalibration. Many of these techniques were developed jointly with successor experiments \citep[\biceptwo\ and the \keckarray;][]{ogburn_2012} 
and they will only grow in importance with improved instrumental sensitivity.

\section{The BICEP1 Instrument}
\label{sec:instrumentsub}

In this section, we summarize the salient features of the \bicep\ instrument. 
More complete details are available in \citet[hereafter \citetalias{takahashi_2010}]{takahashi_2010} and \citetalias{chiang_2010}, as well as in several theses: \citet{yoon_thesis, chiang_thesis, takahashi_thesis, bierman_thesis, moyerman_thesis}.

\bicep\ is a bolometric polarimeter that is specifically designed to search for the signature of inflation in the \bmode\ of the CMB polarization. 
Its detectors, optical path, scan strategy, target region, and site are all designed to provide the highest possible sensitivity while minimizing  polarization systematics. 

The \bicep\ receiver consists of a focal plane of 49 Polarization-Sensitive-Bolometer pairs \citep[PSBs;][]{jones_spie_2003}.
The two detectors in each pair respond to orthogonal linear polarizations.
We derive CMB temperature measurements from the summed pair response and polarization from the differenced pair response.
A two-lens refracting telescope couples the PSBs to the sky, providing full width half-maximum (FWHM)  angular resolution of $0.93^\circ$ and $0.60^\circ$ at 100 and 150~GHz, respectively, and an instantaneous field of view of 18$\deg$. 
The focal plane has 25 PSB pairs at 100~GHz, 22 at 150~GHz, and two at 220~GHz.
These quantities reflect the 2007/2008 configuration. 
For 2006 only, six pairs (three at 100~GHz and three at 150~GHz) were used with Faraday Rotation Modulators \citep[FRMs;][]{moyerman_2013}.
The 220~GHz detectors were introduced in 2007 \citep{bierman_2011}.
 
\bicep\ was installed in the Dark Sector Laboratory ($89\fdg99$~S, $44\fdg65$~W) at the Amundsen-Scott South Pole station to take advantage of the excellent millimeter transparency of the atmosphere above the cold Polar plateau. 
The telescope mount provides three-axis motion: azimuth, elevation, and boresight rotation.
The telescope is fully enclosed inside the warm lab with only the aperture exposed to the polar environment. 
The aperture is surrounded by a co-moving absorptive baffle and a large, fixed reflective ground screen to minimize any potential contamination from warm ground emission.

During its three seasons of operation, \bicep\ observed three fields, concentrating 85\% of its observing time on one CMB region selected for low galactic dust emission. 
This region, called  the ``Southern Hole,'' is located at a right ascension and declination range of  $|\alpha| < 60\deg$ and $-70\deg < \delta < -45\deg$.
The telescope operated on a 48-hour observing cycle, containing four nine-hour ``phases'' targeting the Southern Hole.
Each phase was further divided into ten azimuth-fixed ``scansets,'' approximately 50 minutes long, during which the telescope scanned across the full $60^\circ$ range of azimuth at a fixed elevation.
Each scanset comprises 50 left-going and 50 right-going ``half-scans.''
Each scanset was bracketed by elevation nods, which are small ($1.2^\circ$ peak-to-peak) excursions in telescope elevation used to calibrate relative  detector  gains from the atmospheric signal.
The telescope stepped $0.25^\circ$ in elevation between each scanset and covered the full CMB field after two phases.
The boresight rotation changed between observing cycles, stepping between four orientations (0, 45, 180, and 225$^\circ$) chosen to improve polarization angle coverage.
The remaining 12 hours from each 48-hour cycle were spent on cryogenic service (six hours) and Galactic field observations \citep[six hours; ][]{bierman_2011}.

The focal plane, target field, scan strategy, observation cycle and calibration methods remained unchanged from the 2007 to the 2008 season. 
As a result, for the analysis of the three-year data set, we use the same parameters for the detector transfer functions, relative gains, polarization orientation and efficiency, beam shapes as those presented in \citetalias{chiang_2010}. Similarly, we follow the same procedure for deriving the absolute gain calibration, boresight and detector pointing as in \citetalias{chiang_2010}.

\section{Data Selection}\label{sec:selection}

This analysis uses a data set that has been expanded since \citetalias{chiang_2010}, most significantly by the inclusion of a third year of observations.
The first two years of data include a total of 736 nine-hour CMB phases, with 248 phases in 2006 and 488 phases in 2007; the 2008 season contributes another 270 phases, increasing the total by 37\% to 1006 phases.
The first and last season contribute less integration time due to time spent refining the observing schedule at the start of the 2006 season and time spent on final calibrations at the end of the 2008 season.
As in \citetalias{chiang_2010}, we exclude a small number of incomplete CMB observing phases from the analysis.

\subsection{Observing Efficiency}\label{s:efficiency}
Table \ref{tbl:efficiency} describes \bicep's total observing efficiency, relative to a hypothetical experiment with perfect detector yield, no time spent on cryogenic service or calibration, and no weather cuts.

In the top section of the table, we divide up three calendar years (1095 days) into time spent on summer activities (deployment, upgrades, and the summer calibration described in \citetalias{takahashi_2010}), time spent on CMB observations of the Southern Hole that are used for this analysis, and alternate observing modes, including observations using FRM detectors, published in~\citet{moyerman_2013}, and Galactic field observations, published in~\citet{bierman_2011}.
During its three years of operation, \bicep\ spent 46\% (503 days out of 1095) of its time on the primary science target.

The bottom section of Table \ref{tbl:efficiency} describes the observing cycle efficiency during CMB observations.
Although the final fraction of time spend observing the CMB seem low, this summary of \bicep\ operations describes an instrument that achieved a goal of extremely focused observation on its target field.

\begin{deluxetable}{lcc}
  \tablewidth{\columnwidth}
  \tablecaption{\label{tbl:efficiency} \bicep\ total observing efficiency}
  \tablehead{\colhead{Activity} & \colhead{Days spent\tablenotemark{a}}& \colhead{\S}}

  \startdata

  Installation, upgrades, calibration & 271 days & \ref{sec:selection} \\
  Alternate observing modes & 321 days & \ref{sec:selection} \\
  CMB observations & 503 days & \ref{sec:selection} \\[0.25em]  

  \tableline\\[-0.75em]

  \multicolumn{1}{c}{Down-selection} & Fraction kept in CMB analysis\tablenotemark{b} & \nodata \\[0.25em]

  \tableline\\[-0.75em]

  Cryogenic service and & \multirow{2}{*}{75.0\%}  & \multirow{2}{*}{\ref{s:efficiency}} \\
  \phantom{\hspace{1em}} Galaxy observations \tablenotemark{c}  \\
  Scanset calibrations\tablenotemark{d}  & 83.0\% & \ref{s:efficiency}  \\
  Scan turn-arounds \tablenotemark{e}  & 74.2\%   & \ref{s:efficiency}\\
  Detector yield \tablenotemark{f}  & 89.6\% & \ref{sec:slowtau} \\
  Weather cut  & 92.5\% & \ref{sec:cuts} \\
  Half-scan cuts & 96.8\% & \ref{sec:cuts} \\[0.5em]

  Total pass fraction & 37.1\% & \nodata 

  \enddata

  \tablenotetext{a}{Number of days out of three calendar years, 2006--2008.}
  \tablenotetext{b}{Percentages in this column describe what fraction of the time spent on CMB observations is ultimately included in analysis. The fractions can be applied cumulatively to obtain the total pass fraction (bottom row).}
\tablenotetext{c}{ "Cryogenic service and Galaxy observations"  combines the two six-hour periods from each 48-hour cycle that did not target the CMB field.}
\tablenotetext{d}{"Scanset calibrations" refers to the fraction of time in each observing phase spent performing elevation nods, as well as brief scans over the Galactic field.}
\tablenotetext{e}{"Scan turn-arounds" refer to the periods of acceleration at either end of each azimuthal half-scan, which are cut from the analysis.}
\tablenotetext{f}{Fraction of 100 and 150 GHz detectors used for CMB analysis, weighted average across observing seasons.}

\end{deluxetable}

\subsection{Detector Selection}\label{sec:slowtau}

Of the 49 optically active PSB pairs in the focal plane, the analysis in~\citetalias{chiang_2010} excluded many due to various unexpected behaviors such as poorly behaved transfer functions. 
The transfer functions are used to deconvolve the raw detector timestream into a cleaned timestream usable for analysis and allow us to link the relative gain measured at 0.02~Hz via elevation nods, to the science band (0.1--1~Hz). 
Although we deconvolve the measured transfer functions, a fit to a phenomenological model was used to identify poorly behaved detectors. 
\citetalias{chiang_2010} excluded all detectors with larger than 0.2\% residuals to the model fit.
For this analysis, we include those detectors. 
Although their transfer functions do not follow the common \bicep\ bolometer model, they are measured sufficiently well over the frequency range of interest. 

The analysis of \citetalias{chiang_2010} included 33 PSB pairs (19 at 100~GHz, 14 at 150~GHz) in 2006 and 37 pairs (22 at 100~GHz, 15 at 150~GHz) in 2007.
For this analysis, the count increases to 36 PSB pairs (19 at 100~GHz, 17 at 150~GHz) in 2006 and 43 pairs (23 at 100~GHz, 20 at 150~GHz) in 2007 and 2008.
We exclude the six detector pairs containing Faraday Rotation Modulators for the 2006 season and two 220 GHz pairs from the 2007 and 2008 seasons.

Averaged across the three observing seasons, the number of detector pairs included in the analysis increased by 12\%.
The addition of these detectors  together  with the third season of observations, the total increase in integration time over \citetalias{chiang_2010} is 52\%.

\subsection{Data Cuts}\label{sec:cuts}

Starting from the expanded data set, we remove some data that suffer from bad weather or glitches in the pair-difference timestreams.
The cut criteria are identical to those used in \citetalias{chiang_2010}, but we briefly review them here and present updated cut fractions for the three-year data set.

The first cut is designed to remove entire phases affected by bad weather.
For each PSB and phase, we compute the standard deviation of the ten relative gain measurements from elevation nods made during the phase.
The median of these standard deviations is calculated separately for 100 GHz and 150 GHz PSBs.
If either median value exceeds a threshold, selected to be 20\% of the typical relative gain value, then the entire phase is cut for all detectors at both frequencies.
Applying these criteria reduces the number of phases from 1006 to 930, a cut fraction of 7.5\%.

Next, we cut individual PSB pairs at a single half-scan level according to three criteria:
\begin{itemize}
\setlength{\itemsep}{0em}
\item{A detector pair is cut for all half-scans in a scanset if the A/B relative gain ratio differs by more than 3\% between the two elevation nods bracketing that scanset.}
\item{Pairs are cut for any individual half-scans where the pair-difference data shows significant skewness or kurtosis.}
\item{Half-scans containing large glitches (in excess of 7$\sigma$ from the mean) are cut.}
\end{itemize}
If the combination of these cuts removes more than half of the data for a PSB pair in a particular scanset, 
we take that as evidence of unreliable behavior and cut that pair for the entire scanset.
Altogether, the half-scan cuts remove 3.2\% of the data, significantly less than the weather cut.
The skew/kurtosis cut is the most important of the set; dropping it entirely would lower the cut fraction to 2\%.

\section{Mapmaking}\label{s:mapmaking}

\begin{figure*}[h]
  \resizebox{\textwidth}{!}{\includegraphics{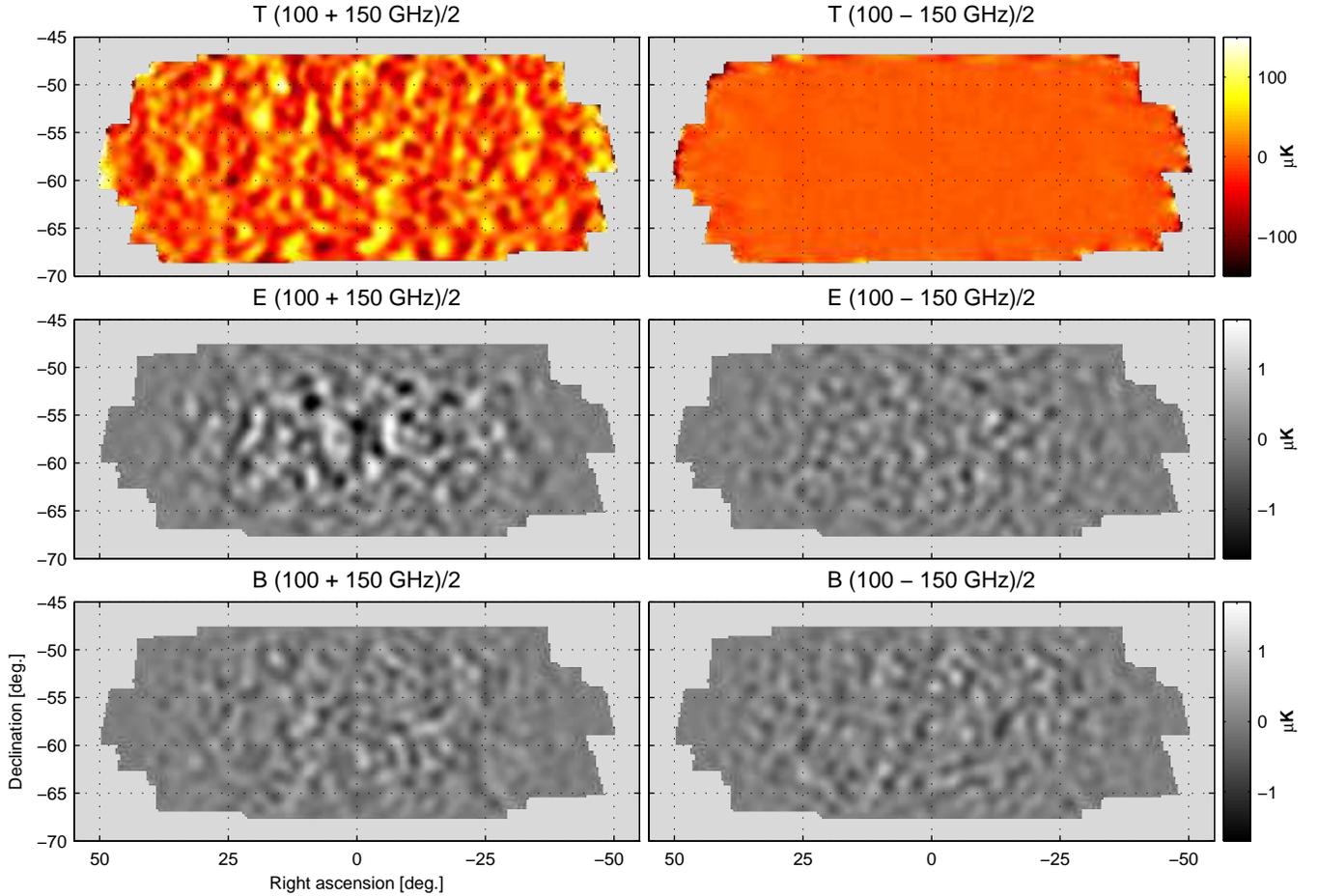}}
  \caption{Maps of CMB temperature and $E/B$-mode polarization generated from three years of \bicep\ observations.
    The 100 and 150~GHz maps are each smoothed to a common $1^\circ$ (full width at half-maximum) beam size before taking the frequency sum (left column) and difference (right column).
    The $E$ and $B$ polarization maps have been additionally filtered to remove power outside the range $30<\ell<200$, in order to emphasize the angular scales of interest for \bicep.
    As in \citetalias{chiang_2010}, \bicep\ detects \emode\ polarization with high significance while the \bmode\ signal is consistent with noise. 
    The $E$- and $B$-mode frequency-difference maps are consistent with noise, indicating that they do not suffer from significant contamination by polarized foregrounds.}
  \label{fig:maps}
\end{figure*}

\citetalias{chiang_2010} presented results from two analysis pipelines. 
The primary result came from an analysis that utilized \healpix\ map binning~\citep{gorski_2005} and the \spice\ power spectrum estimator \citep{chon_2004}.
The ``alternate analysis pipeline'' of \citetalias{chiang_2010}, which uses an equirectangular map pixelization and two-dimensional Fourier Transform for power spectrum estimation, is the only pipeline used for the current work. 
This pipeline has been derived from one originally developed for the \QUAD\ experiment \citep{pryke_2009}.

\subsection{Low-level Processing}\label{s:lowlevel}

The low-level timestream processing is unchanged from \citetalias{chiang_2010}.
Each nine-hour phase is cut to exclude elevation nods bracketing each scanset and periods of acceleration at either end of each half-scan.
The detector transfer function is deconvolved from each detector timestream, which is then low-pass filtered at 5~Hz and down-sampled to 10~Hz. 

A relative-gain correction is calculated for each detector by measuring its response to atmospheric signal during the elevation nods and comparing to the mean response of all good detectors.
After all of the detectors have been scaled to have the same response to the gradient of the airmass, 
the timestreams for each pair are summed and differenced to produce timestreams sensitive to temperature and polarization, respectively.
At this point, we calculate the pointing and polarization angle of each pair and apply a polarization efficiency correction to the pair-difference data.
For details of the pointing and polarization corrections, see \citetalias{takahashi_2010}.

\subsection{Timestream Filtering}\label{s:filtering}
  
Pair-sum and pair-difference timestreams are filtered to remove atmospheric and azimuth-fixed noise prior to map making.
First, we fit and remove a third-order polynomial in azimuth from each half-scan.
Next, for each pair-sum or pair-difference timestream, all half-scans in a scanset are binned together in azimuth and the resulting scan-synchronous structure is subtracted from the data.
This azimuth-fixed filter targets contamination signals from the ground, which remain stationary over the scanset duration while sky signals rotate under sidereal motion.
For discussion of how the filtering affects the signal power measured in the maps and the power-spectrum error bars, see \S\ref{s:bpwf}.

\subsection{Weighting and Map Binning}\label{s:weighting}

The filtered timestreams are binned into maps using the calculated pointing trajectories for each detector pair.
Pair-sum timestreams contribute to a CMB temperature ($T$) map while pair-difference timestreams contribute to maps of the Stokes $Q$ and $U$ parameters, where the particular linear combination of $Q$ and $U$ measured in each sample is calculated based on the measured polarization angle of the detector pair and the telescope boresight orientation.
The $T$, $Q$, and $U$ maps all use a common equirectangular pixelization, with $0.25^\circ$ square pixels at declination $-57.5^\circ$.

In the map binning operation, we weight each half-scan by its inverse variance, calculated as the variance of
the time-ordered data after application of the polynomial and
azimuth-fixed filters. We weight the pair-sum and pair-difference timestream separately.
This choice of weighting is different from the one used in \citetalias{chiang_2010}, which applied a uniform weighting for all half-scans for a particular detector pair in a scanset based on the power spectral density in the range 0.5--1 Hz.
We build weight maps corresponding to the $T$, $Q$, and $U$ maps, which are used as apodization masks for power-spectrum estimation (see \S\ref{s:ps_estimate}).

\subsection{Relative-Gain Deprojection}\label{s:relgain}

As described in \citetalias{takahashi_2010}, a miscalibration in the relative gain of the two detectors in a pair has the effect of leaking the CMB temperature signal into the polarization maps constructed from that detector pair.
Relative-gain mismatch can arise from a difference in spectral response between detectors within a pair.
The atmospheric signal used to measure the relative gains has a different spectrum than the CMB, in the presence of small mismatches in the bandpass within a polarization pair, the relative gains chosen to reject an unpolarized atmospheric signal do not perfectly remove the CMB temperature.

To mitigate systematic uncertainties arising from relative-gain mismatch, we implement a ``relative gain deprojection'' analysis, described in detail in \citet{aikin_2013}.
Over each nine-hour phase, a template timestream for the CMB temperature signal is constructed from the \wmap\ seven-year V-band map \citep{jarosik_2011}, which has been smoothed to the resolution of \bicep.
This template undergoes the same filtering operations as the real data.
Then, we perform a linear regression of the nine-hour pair-difference timestream against the leakage template.
The individual half-scans are weighted by their inverse variance for this regression, the same weighting that is used for map making.
The coefficient obtained from the regression is proportional to  the amplitude of the CMB temperature signal that has leaked into the pair-difference data due to relative gain mismatch.
This temperature-to-polarization leakage signal is subtracted from the data before map making.

While the regression coefficient for a particular detector pair in a single nine-hour phase is noisy, we can average over many phases to obtain a better measurement of the relative gain mismatch for each pair.
We find six detector pairs at 150~GHz and one pair at 100~GHz whose relative gain regression coefficient exceeds 1\%, which is consistent with the findings of \citetalias{takahashi_2010}.
With the relative gain leakage thus removed, we find that the residual from this systematic has been suppressed to negligible levels (see ~\S\ref{s:systematics:relgain}).

In addition to relative-gain deprojection, \citet{aikin_2013} also describes procedures for  deprojection of other differential beam systematics within each polarization pair (differential pointing, beam width, and  beam ellipticity) that cause temperature-to-polarization leakage.
For this analysis, we deproject only the leakage from relative gain mismatch, because it was demonstrated in \citetalias{takahashi_2010} to be the most significant systematic for \bicep.

\subsection{Map Results}\label{s:map_results}

Applying the mapmaking algorithm described above to the three-year dataset yields $T$, $Q$, and $U$ maps at 100 and 150~GHz (available in the data release), which are qualitatively similar to those in Figure~2 of \citetalias{chiang_2010}.
We compare the noise level in these maps, estimated from noise simulations, to that of the two-year maps from \citetalias{chiang_2010}.
When using the same central region, which encompasses 27\% (19.2\%) or 305 (203) square degrees of the 100 (150)~GHz maps, we find that the rms noise is 0.68 (0.50)~$\mu$K-degree.
This is consistent with the 52\% increase in integration time found in \S\ref{sec:slowtau}.

An alternate, and perhaps more natural, measure of map depth is obtained by calculating the weighted standard deviation of a jackknife map (see \S\ref{s:jacktypes}), using the weight map described in \S\ref{s:weighting}.
The jackknife map, chosen to be the season-split temporal jackknife, has the same noise level as the three-year map, but no signal.
The advantages of this technique are that it is tightly related to the procedure used for power spectrum estimation, which uses the same weighting, and that there are no tunable parameters in the procedure (the first map depth calculation requires either a choice of a particular map area or else a threshold on the integration time per pixel).
By this method, we find the noise level in the \bicep\ three-year polarization maps to be 0.90 (0.73)~$\mu$K-degree over an effective area of 446 (291)~square degrees for 100 (150)~GHz.
The noise levels quoted by this method are higher than those listed above because they are determined from a larger fraction of the maps, including pixels near the edge of the field with less integration time.

The 100 and 150~GHz maps are calibrated to common units ($\mukcmb$) and combined to produce the $T$, $E$, and $B$ maps shown in Figure~\ref{fig:maps}.
The left column shows frequency-sum maps and the right column frequency-difference jackknife maps.
The stark contrast between the frequency-sum and frequency-difference temperature maps demonstrates that the CMB temperature anisotropy is observed at high signal-to-noise.
For visual clarity, the $E$ and $B$ maps have been filtered in Fourier space to include only power in the range $30<\ell<200$.
The \emode\ frequency-sum map shows the expected signal while the frequency-difference map is consistent with noise, a confirmation that polarized foregrounds are not detected.
Based on noise and signal simulations (\S\ref{s:nsim} and \ref{s:ssim}), we find that the signal-to-noise ratio in the \emode\ map exceeds unity for the first four bandpowers (up to $\ell \sim 160$) and peaks at a value of $3.3$ for the $90<\ell<125$ band.
The \bmode\ frequency-sum and frequency-difference maps are both consistent with noise.

\section{Power-Spectrum Analysis}\label{s:ps_estimation}

The angular power spectra of the CMB are estimated from two-dimensional  Fourier Transforms (2D FT) of the temperature and polarization maps following the technique described in \citet{pryke_2009}.
Specifically, we make estimates of $\mathcal{D}_b^{XY}$, a binned version of $\mathcal{D}_\ell^{XY} = \ell (\ell+1) \mathcal{C}_\ell^{XY} / 2\pi$.
The indices $X$ and $Y$ denote the two maps used to calculate a particular power spectrum, either the 100 or 150~GHz $T$, $E$, or $B$ maps.
This yields a total of 21 separate power spectra---three spectra for each of $TT$, $EE$, and $BB$; four spectra for each of $TE$, $TB$, and $EB$.

Power measured directly from the maps, $\tilde{\mathcal{D}}_b^{XY}$, can be related to the true CMB power spectra as
\begin{equation}
  \tilde{\mathcal{D}}_b^{XY} = F_b^{XY} \mathcal{D}_b^{XY} + \tilde{N}_b^{XY}.
  \label{eq:psmodel}
\end{equation}
The suppression factor, $F_b^{XY}$, accounts for power removed from the maps by filtering, including relative gain deprojection, as well as smoothing of small scale power due to the 
instrumental beam. Instrumental noise in the maps introduces an additive noise bias, $\tilde{N}_b^{XY}$, to the observed power spectra.

We use a simulation-based technique to derive $F_b^{XY}$ and $\tilde{N}_b^{XY}$, and to ultimately solve for the CMB power spectra.
Full timestream simulations of instrumental noise and/or cosmological signal are processed identically to the real data, including filtering, map making, and power-spectrum estimation.
We make these simulations as realistic as possible, to fully and transparently account for the effect of our analysis pipeline on the data.

\subsection{From Maps to Power Spectra}\label{s:ps_estimate}

The 2D FT is applied to the temperature and polarization (Stokes $Q$ and $U$) maps after they have been apodized by the weight maps.
Because the sky coverage is slightly different for the $Q$ and $U$ maps, the inverse of the mean of the $Q$ and $U$ variance in each pixel is used as a common apodization for those Fourier transforms.

The coordinates of points in the Fourier space maps are $\ell_x$ and $\ell_y$, which represent modes with wave vector in the direction of right ascension and declination, respectively.
The transformed $Q$ and $U$ maps can be rotated into maps representing the even-parity {\emode}s and odd-parity {\bmode}s,
\begin{eqnarray}
  E (\ell_x,\ell_y) & = & + Q (\ell_x,\ell_y) \cos 2\phi + U (\ell_x,\ell_y) \sin 2\phi \\
  B (\ell_x,\ell_y) & = & - Q (\ell_x,\ell_y) \sin 2\phi + U (\ell_x,\ell_y) \cos 2\phi,
\end{eqnarray}
where $\phi = \arctan(\ell_y / \ell_x)$.

After transforming from $Q$ and $U$ to $E$ and $B$, the power spectra, $\tilde{\mathcal{D}}_b^{XY}$, are calculated by multiplying Fourier map $X$ with the complex conjugate of Fourier map $Y$.
This product is scaled by $\ell(\ell+1)/2\pi$, where $\ell = \sqrt{\ell_x^2 + \ell_y^2}$, and then averaged in annular bins.
We report the \bicep\ results in nine bins of uniform width $\Delta\ell=35$, with the first bin spanning $20 \le \ell < 55$ and the ninth bin spanning $300 \le \ell < 335$.

\subsection{Noise Simulations}\label{s:nsim}

To recover the underlying true power spectra, the first step consists of subtracting the noise bias, $\tilde{N}_b^{XY}$.
We form a noise model based on the correlations between detectors in the real data, generate noise-only simulated timestreams, and process them through the same timestream filtering, map making, and power spectrum estimation as the real data.
The resulting simulated noise spectra are then averaged over many realizations to estimate the noise bias.

In \citetalias{chiang_2010}, the noise covariance matrix was based on the correlations between filtered pair-sum and pair-difference timestreams, accumulated over all half-scans in a scanset.
For this analysis, we have chosen instead to use the noise model described in \S5.3 of \citet{pryke_2009}, which calculates a noise covariance matrix for unfiltered individual detector timestreams over an uninterrupted scanset, including the scan turn-arounds that are ultimately cut from both the real and simulated timestreams.
This noise model more closely follows the analysis philosophy of faithfully simulating the real data and then treating the simulated and real timestreams symmetrically, including the filtering step.

Additionally, by deriving the noise model from a full scanset length timestream, instead of a large number of half-scan length segments, we capture the low-frequency atmospheric noise, which persists over many half-scans and is heavily correlated between detectors.
A direct comparison of the noise model used in this analysis with that of \citetalias{chiang_2010}\ is presented in \S\ref{s:consistency_noise}.

We generate 499 independent realizations of \bicep\ noise.
Each realization consists of timestream data for all detectors across all three years of observation.
These simulated timestreams pass through the low-level processing and filtering operations, are binned into maps, and reduced to power spectra.
The noise bias is calculated simply as the ensemble average of the spectra from these 499 noise-only simulations.
While the noise bias is generally close to zero for cross-spectra, those terms are still calculated and subtracted.

\subsection{Signal Simulations}\label{s:ssim}

We generate two classes of signal-only simulations, used to characterize the response of the analysis to CMB signals.
The first set of signal-only simulations, hereafter referred to as ``$E$-no-$B$'', use input CMB spectra generated by \camb\ \citep{lewis_2000} based on \wmap\ five-year best-fit cosmological parameters\footnote{The difference between \wmap\ five-year cosmology and updated parameters from \wmap\ nine-year or \planck\ is negligible for the purposes of calculating the suppression factors and signal variance.} \citep{komatsu_2009}.
As the name implies, these theoretical power spectra have \emode\ power but no {\bmode}s.
For the second set of signal-only simulations, hereafter referred to as ``$B$-no-$E$'', we include primordial tensor perturbations corresponding to $r=0.1$, but explicitly null the $EE$ (and $TE$) power.
The $E$-no-$B$ simulations primarily characterize the leakage between the $E$ and $B$ polarization maps induced by our pipeline, while the $B$-no-$E$ show our pipeline's effect on an input $BB$ signal.

Realizations of these signal simulations are generated from the theory spectra using the \synfast\ utility included in the \healpix\ package\footnote{\url{http://healpix.sourceforge.net}}.
At this stage, the CMB signal is smoothed by convolution with a Gaussian beam; for each simulated sky, two sets of $T$, $Q$, and $U$ maps are produced, corresponding to \bicep\ 100 GHz and 150 GHz beam widths.
The resolution of the \healpix\ maps is $0.11^\circ$ (Nside = 512).
Using the pointing data from the telescope, we simulate timestreams for every detector by performing a second order Taylor expansion interpolation from the nearest pixel center to the actual detector pointing.
Signal-only timestreams are processed through the analysis pipeline, passing through the same filtering and map-making steps as the real data.
For each of the two classes of signal simulation, we apply this process to 499 independent sky realizations drawn from the same underlying CMB power spectra to produce sets of simulated signal-only maps.

When we calculate the power spectra of the $E$-no-$B$ maps, we measure non-zero $BB$ spectra due to $E \rightarrow B$ leakage due to limited sky coverage and filtering.
The ensemble average of $BB$ spectra recovered from the $E$-no-$B$ simulations is subtracted to debias the measured $BB$ spectrum, exactly analogous to the noise bias, but for $BB$ spectra only.
The amplitude of the leakage bias corresponds to $\mathcal{D}_b^{BB} \sim 0.02~{\mu}K^2$ for $\ell \sim 100$.
This value appears significantly larger than the $E \rightarrow B$ leakage reported in \citetalias{chiang_2010} because the \spice\ estimator includes an analytic debiasing operation for the sky cut effect; the level of the leakage quoted in \citetalias{chiang_2010} only accounts for the residual after this analytic debiasing step.
For the 2D FT estimator, the $E \rightarrow B$ leakage is entirely measured from signal simulations.
After debiasing, only the sample variance of the $E \rightarrow B$ leakage signal is important, as this can contribute additional uncertainty to the $BB$ bandpowers\footnote{The excess variance from $E \rightarrow B$ leakage can be reduced through the use of improved estimators \citep[e.g.][]{smith_2006}}.
For this analysis, the leakage contribution to $BB$ bandpower uncertainty is $4 \times 10^{-3}~\muK^2$ (see Figure~\ref{fig:total_error}), which is similar to the value obtained using \spice\ in \citetalias{chiang_2010} and subdominant to instrumental noise.

The power spectra of the $E$-no-$B$ and $B$-no-$E$ signal-only maps are used to correct the bandpower window functions and compute the suppression factors, $F_b^{XY}$, as described in \S\ref{s:bpwf}.
Additionally, we can combine the signal-only maps with noise-only maps to create complete simulations of the real \bicep\ data.
The combination of $E$-no-$B$ signal simulations plus noise are used to derive bandpower error bars, as described in \S\ref{s:ps_results}.
By further addition of scaled versions of the $B$-no-$E$ maps to the $E$-no-$B$ signal and noise maps, we can construct map simulations of a cosmology containing inflationary {\bmode}s at arbitrary values of $r$; these simulations are used in \S\ref{s:t2s} to derive a constraint on $r$ from our data.
Finally, the signal simulation infrastructure is capable of introducing a wide variety of instrumental systematics to the simulated data, such as temperature-to-polarization leakage due to mismatched relative gain or beam imperfections.
These systematic uncertainties are not included in the fiducial set of 499 simulations, but they are included in alternate simulations to characterize power-spectrum systematic uncertainties, in both \S\ref{s:systematics} of this paper and \citetalias{takahashi_2010}.

\subsection{Bandpower Window Functions and Suppression Factors}\label{s:bpwf}

\begin{figure}[t]
  \resizebox{\columnwidth}{!}{\includegraphics{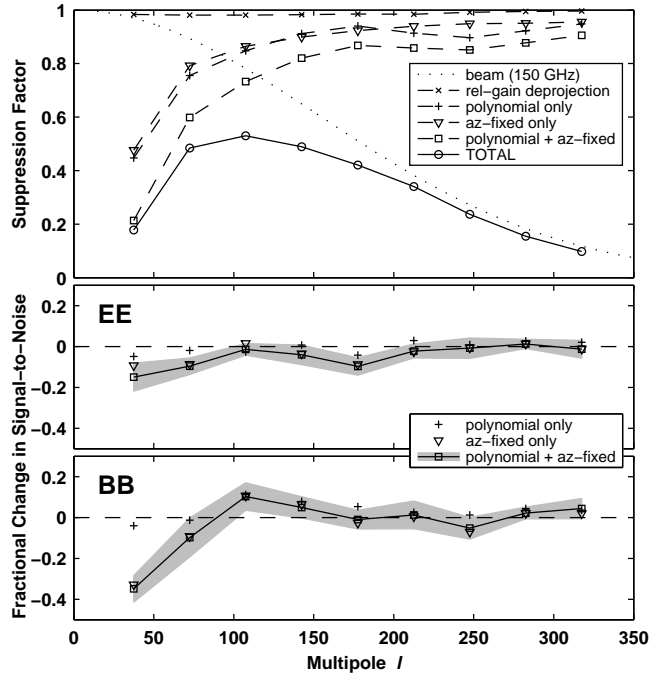}}
  \caption{Shown in the top panel is the suppression factor, $F_b^{150E \times 150E}$, including contributions from 
beam smoothing and our filtering choices. 
At large angular scales, power suppression is dominated by our conservative choice of polynomial and 
azimuth-fixed filters, shown both individually (plus sign and downward triangles) and combined (squares). 
At small angular scales, beam smoothing (dotted line; calculated analytically) dominates. 
The total suppression factor (circles) also includes small contributions from relative gain deprojection
and the pixel window function.
The suppression factor should not be mistaken for a measure of the low $\ell$ performance of the experiment (see text).
The bottom two panels show the actual impact of our filtering choices on the signal-to-noise of $EE$ and $BB$ bandpowers.
The total sensitivity loss from polynomial plus azimuth-fixed filtering is shown by squares, 
with the gray shaded region indicating the $1\sigma$ uncertainty on this calculation. 
The plus signs and downward triangles indicate the loss of sensitivity from polynomial or azimuth-fixed filtering individually. 
While the two filters suppress similar amounts of power in the maps (top panel), only the azimuth-fixed filter has a significant effect on signal-to-noise (bottom panels).}
  \label{fig:fl}
\end{figure}

Bandpower window functions \citep{knox_1999} are used to relate theoretical input spectra, $\mathcal{D}_\ell^{XY}$, to expectation values for the bandpowers measured by \bicep,
\begin{equation}
  \left\langle \mathcal{D}_b^{XY} \right\rangle = \sum_\ell w_{b,\ell}^{XY} \mathcal{D}_\ell^{XY}.
\end{equation}
The window functions, $w_{b,\ell}^{XY}$, are defined to have unit integral over $\ell$.
The shape of the window function is primarily determined by the apodization mask, which mixes power between angular scales.
In Fourier space, this effect can be understood as a convolution of the sharp-edged annulus used by the power spectrum estimator with a smoothing kernel given by the Fourier transform of the mask.
The window functions are also modified by timestream filtering and smoothing of the sky signal by the beam and the map pixelization\footnote{The contribution of pixelation to the window functions and suppression factor is very small for our maps, so we will ignore it for the following discussion, although it is accounted for in our analysis.}. 
This is most significant for the first bin, as the timestream filtering preferentially removes power from the largest angular scales in that range.

To calculate the bandpower window functions, we start by calculating the window function corresponding only to the apodization mask, $m_{b,\ell}^X$, following the procedure described in \citet{challinor_2005}.
To better capture variations in the suppression factor, we calculate this ``mask window function'' for $\ell$-bins, $b'$, that are smaller than the usual bins by a factor of four (annuli in the Fourier plane with width $\Delta\ell = 8.75$).
Additionally, we define these bins across a much wider range of angular scales, from the origin of the Fourier plane out to $\ell \sim 500$, to capture the leakage of power between angular scales.
The mask window function and signal-only simulations are used to make a preliminary estimate of the suppression factor for the fine angular bins,
\begin{equation}
  F_{b'}^{XY} = \frac{\left\langle \tilde{\mathcal{D}}_{b'}^{XY} \right\rangle}{\sum_\ell m_{b',\ell}^{XY} \mathcal{D}_\ell^{XY}}.
  \label{eq:supfac_est}
\end{equation}
Here, the expectation value in the numerator is an ensemble average over power spectra calculated in fine angular bins from the simulated signal-only maps, while the term in the denominator comes from applying the mask window function to the input spectrum used to generate those simulations. 
The suppression factor describes how the telescope and analysis pipeline remove power at each angular scale through beam smoothing and timestream filtering.
The $E$-no-$B$ simulations are used to make this calculation for the $TT$ and $EE$ suppression factors.
The $BB$ suppression factors are calculated using $B$-no-$E$ simulations.
For the $TE$, $TB$, and $EB$ cross-spectra, we use the geometric mean of the suppression factors for the two corresponding auto-spectra (\textit{e.g.} the $TE$ suppression factor is the geometric mean of the $TT$ and $EE$ suppression factors).

Next, the bandpower window functions are corrected by multiplying the mask window functions with a smoothly interpolated version of the suppression factor.
After renormalizing the bandpower window functions, this procedure can be iterated, with the more accurate bandpower window functions substituted in place of the initial mask window functions in Equation~\ref{eq:supfac_est}.
In practice, this procedure converges very quickly; we perform three iterations, but essentially all of the modification to the window functions occurs in the first iteration.

\begin{figure}[t]
  \resizebox{\columnwidth}{!}{\includegraphics{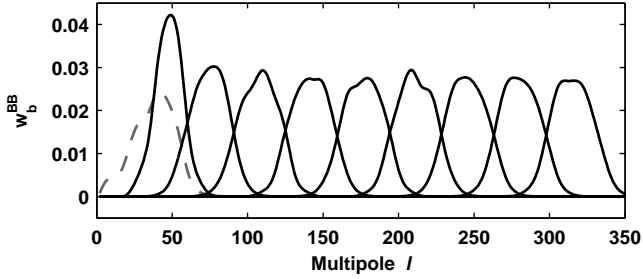}}
  \caption{\bicep\ bandpower window functions for the 150 GHz $BB$ auto-spectrum. The gray dashed line shows the mask window function for the first $\ell$-bin. Bandpower window functions for other spectra are visually similar.}
  \label{fig:bpwf}
\end{figure}

At the end of this procedure we have final suppression factors and corresponding corrected bandpower window functions. 
The finely binned bandpower window functions and suppression factors are then merged to recover equivalent functions appropriate for the nine $\Delta\ell=35$ bins used to report \bicep\ results.
The top panel of Figure \ref{fig:fl} shows the suppression factor for the $EE$ spectrum, as well as individual contributions from polynomial and azimuth-fixed filtering, relative-gain deprojection, and beam smoothing.
Figure \ref{fig:bpwf} shows the bandpower window functions for $BB$ only\footnote{The other window functions are calculated and included in the \bicep\ data release, but are not included in Figure \ref{fig:bpwf}.}.
For the first $\ell$-bin, we also plot the mask window function, to highlight the change in that bin due to filtering.

We emphasize that the power suppression due to filtering is not a measure of the loss of information; signal and noise are both suppressed by filtering.
The suppression factor is also not an indicator of the noise spectrum; for our pipeline the filtering choices are set by hand and have been
chosen conservatively (see below).  The suppression factor is not a meaningful measure of low $\ell$ performance.

To measure the true impact of filtering, we run simulations with the filtering relaxed by either reducing the order of the polynomial filter from three to one, turning off the azimuth-fixed filter, or both.
The fractional change in signal-to-noise (bottom panels of Figure~\ref{fig:fl}) is calculated relative to a ``minimal filtering'' analysis that uses a first-order polynomial filter and no azimuth-fixed filtering.
This statistic depends on both the signal-to-noise regime and the assumed shape of the signal spectra. 
Uncertainty in the statistic, due to the finite number of realizations, is estimated based on the variance between subsets of the realizations.
For $EE$, the combination of polynomial and azimuth-fixed filtering reduces the first bin sensitivity by 15\%.
For $BB$, the first bin sensitivity is reduced by 35\%. 
This factor is larger for $BB$ than $EE$ because the $BB$ spectrum is noise dominated and because the $EE$ spectrum rises steeply across the first bin $\ell$ range, so loss of information from the lowest multipoles is less important.

Comparing the loss of sensitivity for the combined polynomial and azimuth-fixed filtering (squares) with the similar factors for polynomial (plus signs) or azimuth-fixed (triangles) filtering only, we find that nearly all of the information loss in the first bin is due to the azimuth-fixed filtering.
The azimuth-fixed filtering targets ground-fixed or scan synchronous contamination but also attenuates long wavelength sky signals.
The scale at which the azimuth-fixed filter affects bandpower sensitivity is a direct consequence of the choice of timescale 
(one scanset, $\sim50$ minutes) used to construct the azimuth-fixed template.
Polynomial filtering affects signal and noise more equally.  Third-order filtering was chosen based on our temperature data.  
As seen in Figure~\ref{fig:fl} (top panel, triangles), relaxing the polynomial filtering to first order for our polarization data
would change the suppression factor dramatically at low $\ell$, 
but brings no significant benefit in signal-to-noise (bottom panels),
so for simplicity we retain the same filtering choices for both temperature and polarization data.

\subsection{Power-Spectrum Results} \label{s:ps_results}

After calculating $\tilde{N}_b^{XY}$ and $F_b^{XY}$, we can solve for the underlying CMB power spectra, $\mathcal{D}_b^{XY}$, using Equation~\ref{eq:psmodel}.
As discussed in \S\ref{s:ssim}, the average $BB$ signal from $E$-no-$B$ signal simulations is subtracted from the $BB$ bandpowers along with the noise bias.

\begin{figure*}[h]
\begin{center}
\resizebox{0.95\textwidth}{!}{\includegraphics{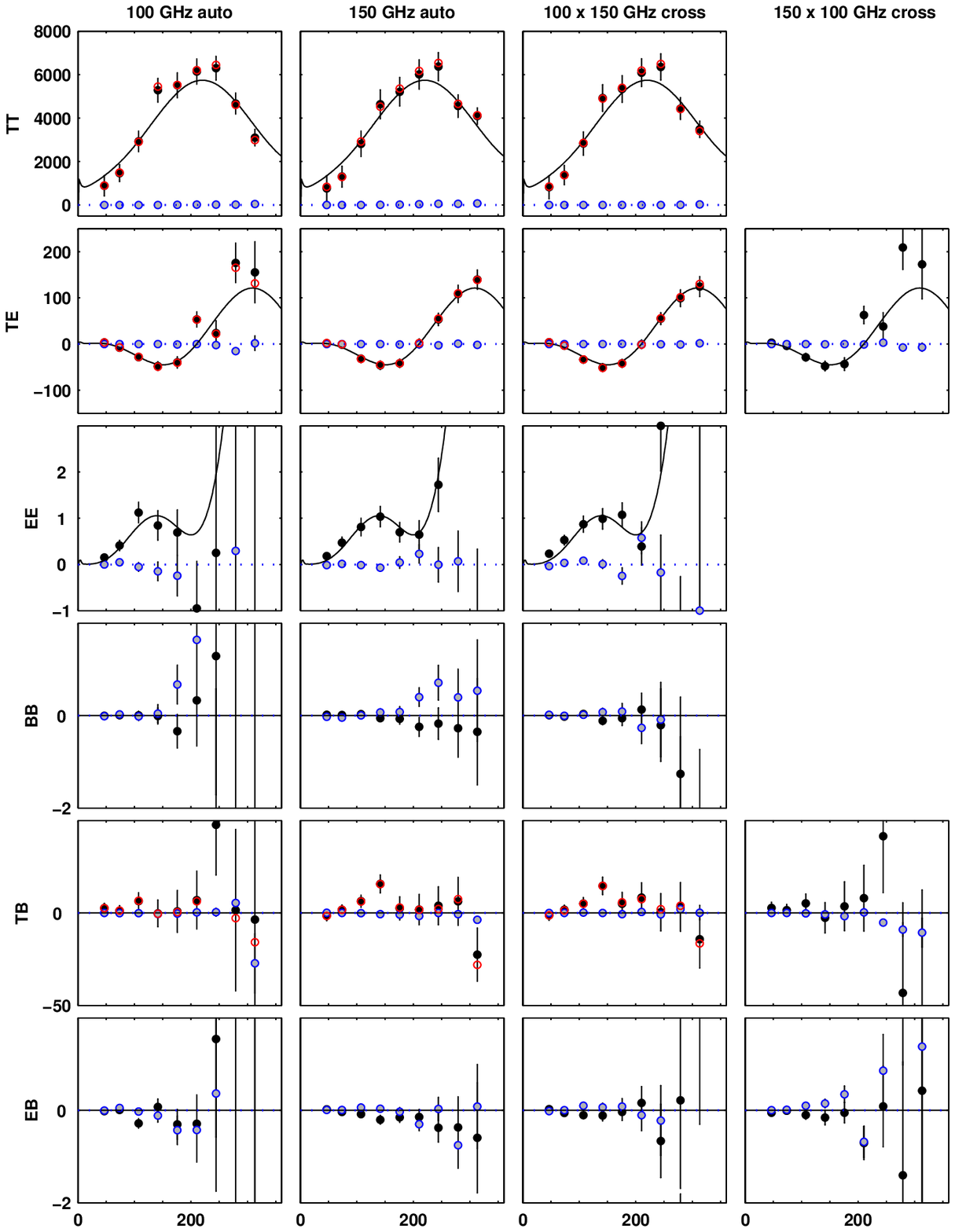}}
\caption{\bicep\ individual-frequency CMB power spectra.
The horizontal axis is multipole $\ell$, and the vertical axis is $\ell(\ell+1)\mathcal{C}_\ell/2\pi$ in 
units of $\mukcmb$. Black points show the full set of \bicep\ power spectra up to $\ell = 350$ with statistical error bars (including sample variance) only. The spectra agree well with a \lcdm\ model (black lines) derived from \wmap\ five-year data and $r = 0$.
The blue points correspond to the boresight-angle jackknife. The red open circles show the $TT$, $TE$ and $TB$ spectra calculated using one half of the \bicep\ boresight-angle 
jackknife maps as the temperature map, as described in \S\ref{s:jacktypes}.}
\label{fig:all_spectra}
\end{center}
\end{figure*}

By combining the noise-only maps with the $E$-no-$B$ signal-only maps, we create a set of 499 signal-plus-noise maps; each of these maps represents a full simulation of the three-year \bicep\ observations with independent noise and signal realizations.
We calculate the power spectra of the signal-plus-noise maps and process them identically to the real data, subtracting the noise bias and dividing by the suppression factor.
This set of simulated power spectra is used to directly determine the covariance matrix of the bandpowers.
All 21 \bicep\ power spectra are shown in Figure \ref{fig:all_spectra}, with error bars given by the square root of the diagonal elements of the bandpower covariance matrix.

\begin{figure*}
\begin{center}
\resizebox{\textwidth}{!}{\includegraphics{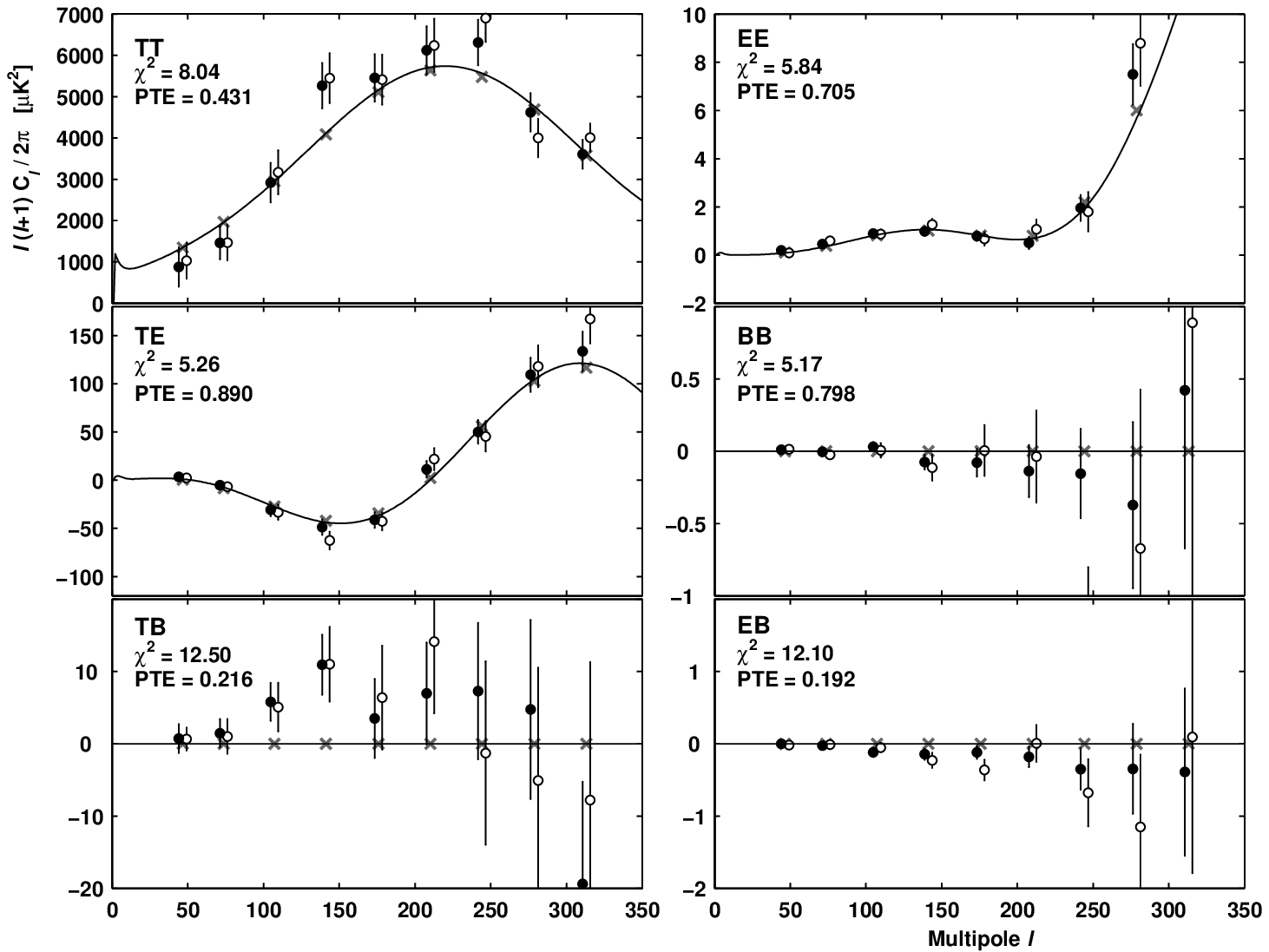}}
\caption{\bicep\ frequency-combined power spectra (black points) are in excellent agreement with a \lcdm\ model (black lines) derived from \wmap\ five-year data.
The $\chi^2$ (for nine degrees of freedom) and PTE values from a comparison of the data with the model are listed in the plots.
Gray crosses denote bandpower expectation values for the model.
Power spectrum results from \citetalias{chiang_2010} are shown by open circles and are offset in $\ell$ for clarity. In both cases, the error bars are the square root of diagonal elements in the frequency-combined bandpower covariance matrix described in \S\ref{s:ps_results}, and do not include systematic uncertainties.}
\label{fig:data_vs_lcdm}
\end{center}
\end{figure*}

The full statistical power of \bicep\ is realized by combining the frequency auto and cross-spectra.
For each bin of each spectrum, we take a weighted average between the corresponding bandpowers from three (or four) different frequency combinations.
Weights are calculated as
\begin{equation}
  \mathcal{W}_i = \sum_j \mathcal{M}_{ij}^{-1},
  \label{eq:comb_weight}
\end{equation}
where $\mathcal{M}_{ij}$ is the appropriate 3$\times$3 (or 4$\times$4) block of the bandpower covariance matrix.
After performing the same frequency combination on the simulated spectra, we can calculate a new bandpower covariance matrix for the frequency-combined spectra.
The frequency combined power spectra are shown in Figures \ref{fig:data_vs_lcdm} and \ref{fig:eebb}.
For the chosen bin width, $\Delta\ell = 35$, bandpowers in adjacent $\ell$-bins have 8--10\% positive correlations.
Elements of the covariance matrix that are separated by more than two $\ell$-bins are too small to be well measured from our finite sample of simulations, so we set them to zero.

\section{Consistency Tests}

During the analysis that led to \citetalias{chiang_2010}, two different analysis pipelines worked in parallel and compared results systematically up to the final 
result on the tensor-to-scalar ratio, $r$.  Only one of these pipelines was used for the three-year \bicep\ analysis so this side-by-side confirmation is not available.
We have therefore tested that the results are insensitive to both the algorithmic and the data-selection changes between this result and those presented in \citetalias{chiang_2010}. 

In addition, we subject all the spectra to the same set of jackknife null tests as in \citetalias{chiang_2010} to probe for systematic contamination.
The $TE$ and $TB$ jackknives merit special discussion to demonstrate that the non-jackknife spectra are valid despite some formal null-test failures.

\subsection{Consistency with Two-Year Results}\label{s:consistency}

Here, we review all the algorithmic and data-selection differences between this analysis and that of \citetalias{chiang_2010} and present the effect of these changes on the polarization bandpowers.

\subsubsection{Consistency between Analysis Pipelines}

As described in \S\ref{s:mapmaking}, the two separate pipelines used in \citetalias{chiang_2010} differ mainly in map format (\healpix\ versus equirectangular map pixelization) and power 
spectrum estimator (\spice\ vs 2D  FT), so we expect minor differences in their respective maps and derived power spectra. In \citetalias{chiang_2010} we have carefully cross-checked those 
differences and found that they are negligible compared to the statistical uncertainty of the spectra. Specifically, focusing on the frequency-combined $BB$ spectrum (see Figure~11 in \citetalias{chiang_2010}), we found that the bandpowers from the two pipelines used in \citetalias{chiang_2010} agree to better than $1\sigma$ with better than $0.2\sigma$ agreement in the first four bins.

\subsubsection{Noise Model Consistency}\label{s:consistency_noise}

As discussed in \S\ref{s:nsim}, this analysis uses a more sophisticated noise model to better capture the details of instrumental and atmospheric noise.
The noise bias calculated from simulations is subtracted from the bandpowers, so noise misestimation leads directly to bias in the power spectra.
To assess the impact of this change, we compare sets of 100 noise realizations that differ only in the choice of noise model: either the model described in \S\ref{s:nsim} or the one discussed in \S6.1 of \citetalias{chiang_2010}.
We process these two sets as described in \S\ref{s:ps_estimation} and calculate the noise bias for each model, focusing on the $BB$ spectrum for which a change in noise bias would have the most significant impact.
We find that the two calculations differ by at most 0.03 ${\mu}K^2$, with a largest fractional shift of 6\%.
Moreover, the noise power does not increase or decrease uniformly across $\ell$-bins, so the small difference is averaged down further when constraining $r$.
Using the \citetalias{chiang_2010} noise model would change the observed value of $r$ by just 0.03.
As a test of the accuracy of bandpower error bars derived from noise simulations, we recalculate the suite of jackknife null tests described in \S\ref{s:jacktypes}, and find that they are indistinguishable when we use either set of noise simulations.

\subsubsection{Bandpower Window Function Consistency}\label{s:consistency_bpwf}

The bandpower window functions reported in \citetalias{chiang_2010} are derived from the \spice\ kernel, making it difficult to provide a direct comparison to the window functions of the current analysis.
However, the procedure described in \S\ref{s:bpwf}, which modifies the window functions to account for the effects of filtering and beam smoothing, was not applied in \citetalias{chiang_2010}.
This change is significant for the first $\ell$-bin only.
For that bin, the timestream filtering shifts the window function to higher $\ell$ and leads to a smaller value for the suppression factor (\textit{i.e.} increased suppression of power).
The instrumental noise contribution to the bandpowers is scaled by the inverse of the suppression factor, as can be seen from Equation~\ref{eq:psmodel}, so the net result is a larger error bar for the first bin bandpowers.
If we compare the error bars obtained for the $BB$ spectrum using the mask window function in place of the bandpower window function, we find that they are underestimated by 40\% for the first $\ell$-bin only.
This test is only applicable for the comparison between window function treatments for the 2D FT power spectrum estimator; a comparison to the error bars reported in \citetalias{chiang_2010} includes a different power spectrum estimator, which leads to an entirely separate change in the error bar estimate.

Tests of consistency between observed and simulated bandpowers, either for jackknife tests or comparison with \lcdm\ cosmology, are independent of the suppression factor and are not affected by this change.

\subsubsection{Relative Gain Deprojection}\label{s:consistency_deproj}

This analysis incorporates a deprojection technique to remove any temperature leakage due to relative gain mismatch (see \S\ref{s:relgain} and~\S\ref{s:systematics:relgain}) for details). We perform two complete analyses with and without this deprojection and find that the final \BB\ bandpowers shift by less than $0.1\sigma$ except for the bin at $\ell = 107.5 $ which shifts down by $0.6\sigma$. Although this change is small, the application of relative gain deprojection significantly reduces this source of systematic uncertainty.

\subsubsection{Consistency with Inclusion of Additional Data}\label{s:consistency_data}

In this analysis, the inclusion of the third season of data and previously discarded detectors has increased the total integration time by 52\% over \citetalias{chiang_2010}. We include the new data incrementally and use sets of 100 signal-plus-noise realizations to determine whether the resulting shifts in bandpowers are statistically significant. 
We find that re-including the set of detectors with abnormal transfer functions produces negligible shifts in all bandpowers, much smaller than the difference between the two pipelines, for example. 
Including the third season of data produces shifts that are consistent with the expectation from simulations.
As additional confirmation that the third season is consistent with the first two, we calculate a modified version of the season-split temporal jackknife (see \S\ref{s:jacktypes}) where the two jackknife halves are the 2006+2007 seasons and 2008 season. This jackknife test passes, though we do not include it in our standard suite. 

Examining the ratio of the error bars from two simulation sets which differ only by the additional data, we confirm that the sensitivity improvement for the $BB$ spectra is purely proportional to the increase in integration time, matching the expectation for the noise dominated case. 

\subsection{Jackknife Null Tests}\label{s:jacktypes}

We perform ``jackknife'' tests to verify that the power spectra are free of systematic contamination. These jackknives are statistical tests in which the data are split in two halves, processed to form maps, and then the maps are differenced. The power spectra of the differenced, or jackknife, map are tested for consistency with zero power, to within the uncertainty derived from simulations. For this analysis, we adopt the same five jackknife splits that were tested in \citetalias{chiang_2010}: scan direction, elevation coverage, boresight rotation, season-split temporal, and focal plane $QU$ jackknife. They are designed specifically to be probe  for instrumental systematic effects. We did not perform the eight-day temporal jackknife and the season-split temporal jackknife has been updated to split the three seasons into two even halves (except for the test noted in \S\ref{s:consistency_data}). For brevity, we refer the reader to the description of these jackknife tests in \S8.1 of \citetalias{chiang_2010}.

\begin{figure}[t]
\resizebox{\columnwidth}{!}{\includegraphics{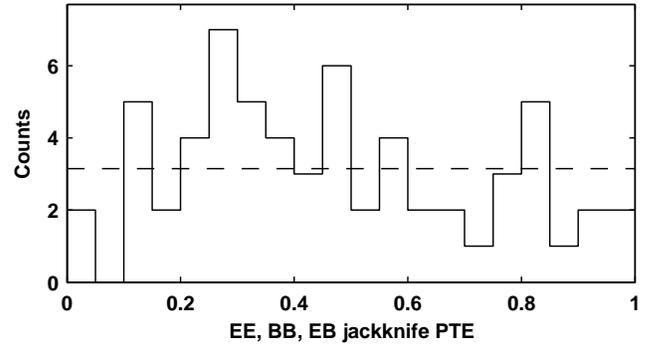}}
\caption{Probabilities-to-exceed from $\chi^2$ tests of 50 polarization-only ($EE$,
$BB$, and $EB$) jackknives are consistent with a uniform
distribution between zero and one (dashed line).}
\label{fig:pte}
\end{figure}

To test the jackknife spectra, we evaluate the $\chi^2$ goodness of fit, with nine degrees-of-freedom, to the null hypothesis. 
Due to differences in filtering and sky coverage between the two halves of a jackknife maps, we expect small levels of residual signal.
We account for this by evaluating the probability-to-exceed (PTE) for the real data against the distribution of $\chi^2$ values from simulations, which should contain the same residuals, rather  
than using a theoretical $\chi^2$ distribution. 
We evaluate the results of the jackknife tests by the following measures:
\begin{itemize}
\setlength{\itemsep}{0em}
\item{The fraction of jackknife spectra with PTE smaller than 5\% should not be significantly larger than 5\%.}
\item{None of jackknife spectra should have a PTE that is excessively small (${\ll}1\%$).}
\item{The PTE from all jackknives should be uniformly distributed between zero to one.}
\end{itemize}
The PTE from jackknife $\chi^2$ tests of all frequency combinations for the polarization-only ($EE$, $BB$, $EB$) spectra are presented in Table~\ref{t:ptes}. Only two of the tests (out of 50) have PTE values less than 5\% and neither one
is exceedingly low (2.6\% for the $150\times100$ $EB$ spectrum from the boresight angle jackknife and 4.2\% for 150~GHz $BB$ season-split jackknife). Figure~\ref{fig:pte} shows that the 
histogram of the PTE values follows the expected uniform distribution.  
The boresight rotation jackknife spectra is plotted in Figure~\ref{fig:all_spectra} (blue points) for comparison with the non-jackknife spectra.

\begin{deluxetable}{c c c c c}
  \tablecaption{Jackknife PTE Values from Polarization-Only $\chi^2$ Tests \label{t:ptes}}
  \tablehead{\colhead{Jackknife} &
    \colhead{100~GHz} & \colhead{150~GHz} & \colhead{100$\times$150} & \colhead{150$\times$100}}
  \startdata
  Scan direction \\
$EE$ & 0.756 & 0.124 & 0.575 \\
$BB$ & 0.244 & 0.246 & 0.327 \\
$EB$ & 0.679 & 0.804 & 0.148 & 0.391 \\
\\
Elevation coverage \\
$EE$ & 0.341 & 0.471 & 0.581 \\
$BB$ & 0.106 & 0.581 & 0.319 \\
$EB$ & 0.335 & 0.639 & 0.273 & 0.764 \\
\\
Boresight angle \\
$EE$ & 0.733 & 0.952 & 0.192 \\
$BB$ & 0.493 & 0.257 & 0.836 \\
$EB$ & 0.489 & 0.251 & 0.104 & 0.026 \\
\\
Season split \\
$EE$ & 0.495 & 0.156 & 0.804 \\
$BB$ & 0.230 & 0.042 & 0.525 \\
$EB$ & 0.471 & 0.421 & 0.918 & 0.898 \\
\\
Focal plane $QU$ \\
$EE$ & 0.986 & 0.411 & 0.383 \\
$BB$ & 0.287 & 0.834 & 0.451 \\
$EB$ & 0.279 & 0.244 & 0.784 & 0.541 \\

  \enddata
\end{deluxetable}

In addition to the $\chi^2$ tests, we compare the jackknife bandpower deviations, defined as the ratio of bandpower values to their error bars, against the simulations. This provides a strong test 
for a coherent bias in the bandpowers, which could be caused by misestimation of the noise bias. For all the polarization jackknife spectra, the sum of bandpower deviations are found to be 
consistent with the simulated distributions.

The jackknife PTE values for those power spectra including the $T$ map ($TT$, $TE$, and $TB$) are not shown in Table~\ref{t:ptes}; they display significant failures. 
The PTE distributions for $TE$  
and $TB$ have an excess of values between $0.05$ and $1{\times}10^{-5}$ and most of the $TT$ jackknives have extremely small PTE values. We hypothesize that the $TE$ and $TB$ failures 
are caused by imperfect signal cancelation in the temperature jackknife maps, as opposed to the polarization maps.

Working from this hypothesis, we have built estimates of $TE$ and $TB$ jackknife contamination derived from the observed $TT$ jackknife failures.
These estimates are indeed consistent with the observed $TE$ and $TB$ jackknife bandpowers.

Additionally, we explore a special type of $TE$ or $TB$ jackknife, referred to as ``half-jackknives,'' where we calculate the cross-spectrum between the jackknife polarization ($E$ or $B$) map and a full, \textit{i.e.} non-jackknife, temperature map. In these half-jackknives, an inconsistency in the $E$ or $B$ map would still show up as a jackknife failure in excess of the simulations while an inconsistency driven by the temperature map would disappear. All 
the PTE for the 20 $TE$ and $TB$ half-jackknife tests exceed 5\%, confirming again that the temperature maps are the likely source of the $TE$ and $TB$ jackknife failures.

As a final check, we confirm that the formal failures in the $TE$ and $TB$ jackknife tests reflect contamination that is at a negligible level compared to the noise level of those non-jackknife spectra. In \citetalias{chiang_2010}, this was done by comparing the \bicep-only $TE$ and $TB$ spectra to spectra constructed using the \wmap\ temperature map. 
Here, we perform an equivalent test by calculating alternate $TT$, $TE$, and $TB$ spectra using just one half of any of the temperature jackknife maps. 
If the jackknife failures are indicating contamination at a level that significantly affects the \bicep\ power spectra, we would see a large difference between power spectra calculated from the 
discrepant halves of the jackknife map.
These alternate spectra are shown for the case of the boresight angle jackknife as red 
open circles in Figure~\ref{fig:all_spectra}. The difference between the two is consistently less than 5\% of the error bars for the $TT$, $TE$, or $TB$ non-jackknifed spectra.

\section{Systematic Uncertainties}\label{s:systematics}

\begin{figure}[t]
\resizebox{\columnwidth}{!}{\includegraphics{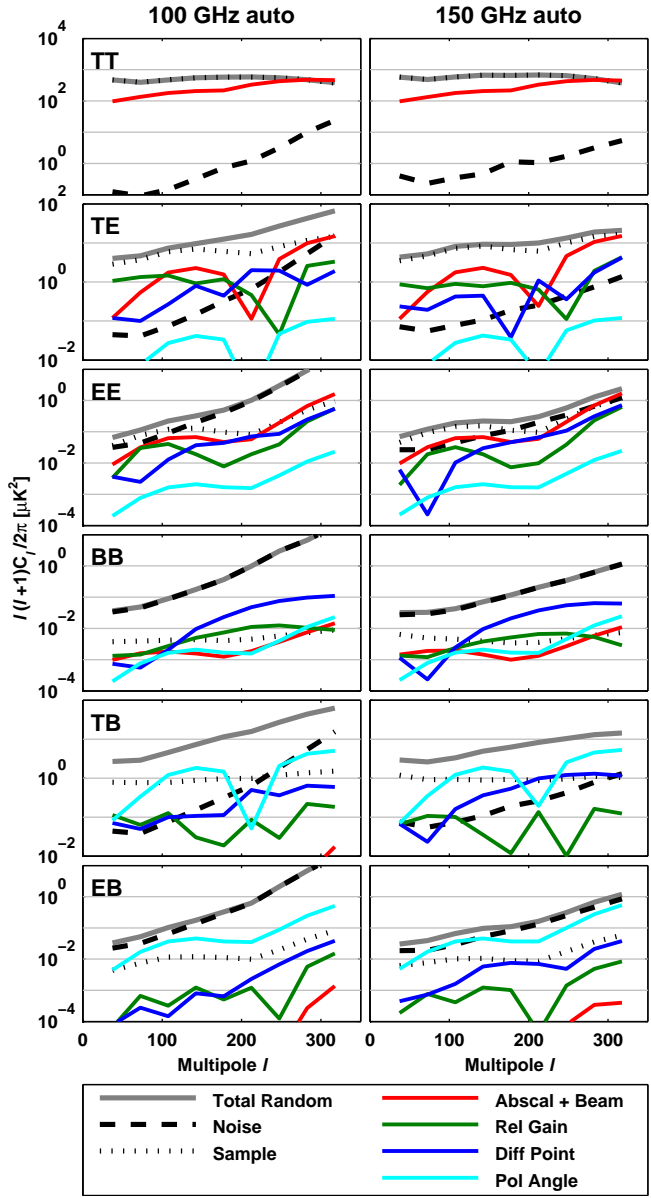}}
\caption{Summary of statistical and systematic uncertainties for the \bicep\ three-year result.
Total random (solid gray line) indicates the overall statistical uncertainty due to both instrumental noise (black dashed) and CMB sample variance (black dotted).
Systematic uncertainty contributions from absolute calibration and beam width (red line, \S\ref{s:abscal}), relative gain mismatch (green line, \S\ref{s:systematics:relgain}), differential pointing (blue line, \S\ref{s:systematics:diffpoint}), and polarization angle calibration (cyan line, \S\ref{s:systematics:polrot}) are also included.
Sample variance dominates the $TT$ and $TE$ spectra.
The $EE$ (and $TB$) spectra are dominated by sample variance at low $\ell$ and noise at high $\ell$.
Noise dominates the $BB$ (and $EB$) uncertainty at all angular scales.
Polarization angle calibration is a significant source of uncertainty for both $TB$ and $EB$.}
\label{fig:total_error}
\end{figure}

For a comprehensive study of systematic uncertainties, we refer the reader to \citetalias{takahashi_2010}.
Here we address only the dominant sources of systematic uncertainties as identified in \citetalias{takahashi_2010} and \citetalias{chiang_2010}, specifically updating the estimates of relative gain and polarization orientation uncertainties.
Figure~\ref{fig:total_error} summarizes the statistical and systematic uncertainties estimates for \bicep\ power spectra.

\subsection{Absolute Calibration and Beam Width Uncertainty}\label{s:abscal}

We follow the same procedure as in \citetalias{chiang_2010} to derive the absolute gains for the three-year maps. Given the similarity of the results, the value and uncertainties of the  absolute-calibration and beam-width  remain unchanged from \citetalias{chiang_2010}. These systematic uncertainties are multiplicative in the observed power spectra; they can lead to incorrect scaling of an observed signal but will not produce false {\bmode}s. The sum of the absolute calibration and beam width systematics are plotted in Figure~\ref{fig:total_error}. For all 
polarization spectra, these uncertainties are a small fraction of the total statistical uncertainty.

\subsection{Relative Gain Mismatch}\label{s:systematics:relgain}

For the \citetalias{chiang_2010} analysis, relative gain mismatch caused by imperfectly differenced detector pairs
was judged the leading source of possible $BB$ contamination, estimated to produce a bias on $r\le 0.17$ \citepalias{takahashi_2010}.
Although this exceeded the $r = 0.1$ benchmark set, it remained small compared to the statistical uncertainty 
of the two-year result. With the addition of a third year of data and additional detectors,
we are motivated to use the relative gain deprojection technique described in \S\ref{s:relgain} \citep[and further detailed in][]{aikin_2013} to keep the potential contamination 
from relative gain leakage well below the statistical uncertainty.

To quantify the remaining leakage after deprojection, we compare three sets of simulated data: the first two include relative gain mismatch at the level observed for \bicep\ and either do, or do not, employ deprojection to remove the resulting temperature-to-polarization leakage. The third set of simulations feature perfectly matched relative gains and does not use relative gain deprojection.

Comparing the simulations with leakage to the idealized case, we find that the excess $EE$ and $BB$ bandpower is greatly reduced by the deprojection technique. Figure~\ref{fig:total_error} includes a residual leakage component that is at least an order of magnitude below the statistical uncertainty for $BB$, and corresponds to a potential bias on $r$ of less than 0.03.

\subsection{Beam Mismatch}\label{s:systematics:diffpoint}

As in \citetalias{chiang_2010}, differential pointing, an offset in the beam centers for paired detectors, is the leading systematic from beam mismatch. Other beam mismatch terms, differential beam ellipticity 
and differential beam width, are far below the $r=0.1$ benchmark established in \citetalias{takahashi_2010}. 
Using simulations that contain the observed \bicep\ differential pointing offsets, we measure the resulting excess bandpower shown in Figure~\ref{fig:total_error}, and confirm that the possible bias from the differential pointing is less than $r=0.02$.
\citet{aikin_2013} includes description of deprojection techniques that can be applied to correct differential pointing, ellipticity, and beam width, but we find these corrections to be unnecessary for \bicep.

\subsection{Polarization Orientation}\label{s:systematics:polrot}

An error in the orientation of detectors can lead to rotation of $E$- into $B$-modes.  The $BB$ spectrum is affected only at second order,
but the $TB$ and $EB$ spectra are more sensitive to such an effect.
In \cite{kaufman_2013} we revisit the systematic uncertainty on
our standard dielectric-sheet based polarization orientation calibration
by comparing it to three alternative calibrations, increasing this
uncertainty from the previously reported 0.7$\deg$ (\citetalias{takahashi_2010}) to 1.3$\deg$.
Figure~\ref{fig:total_error} shows the systematic uncertainty on bandpowers due to the updated calibration error.

The $TB$ and $EB$ spectra can also be used to ``self-calibrate'' the
polarization angle from the CMB itself  \citep{keating_2013}, also as described in \cite{kaufman_2013}.
Applying each of these four alternative calibrations produces small
shifts in the $BB$ spectra.  The maximum shift among these cases
in the estimate of $r$ is less than $0.04$.  The ``self-calibrated'' case produces a 95\% upper limit $r < 0.65$ (vs. $r < 0.70$).
We consider these shifts small for all results based on the $BB$ spectra and therefore keep the original estimate of polarization orientation angle for the main result of this paper.

\section{Foregrounds}\label{s:foregrounds}

In \citetalias{chiang_2010}, we estimated the level of foreground contamination in \bicep\ maps and found it to be negligible.
Here we update those estimates for the three-year analysis and with recently available foreground models.
Given the modest improvement in sensitivity, we expect polarized foregrounds to remain undetected.
Nevertheless, we provide new upper limits on possible contamination from Galactic diffuse emission and compact sources.
As further evidence that our spectra are free of significant contamination, we present a $100-150$ GHz frequency jackknife.

\begin{figure}[t]
  \resizebox{\columnwidth}{!}{\includegraphics{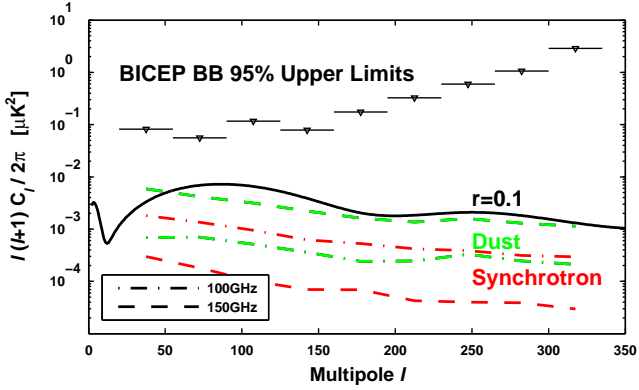}}
  \caption{
    The estimated Galactic dust and synchrotron emission in the \bicep\ field is well below current $BB$ upper limits.
    These foreground emission estimates come from processing the Planck Sky Model foreground maps \citep{delabrouille_2013} through the \bicep\ pipeline. The $BB$ upper limits are derived from the bandpower likelihoods calculated in \S\ref{s:bandpower_likelihood}.}
  \label{fig:galactic_foreground}
\end{figure}

To estimate the effects of Galactic diffuse emission, we use simulated Planck Sky Model \citep[][PSM v. 1.7.7]{delabrouille_2013} polarization maps including thermal dust and synchrotron emission.
We process these maps through the \bicep\ pipeline to estimate the contamination in our field.
The result is more than an order of magnitude smaller than our upper limits on the $BB$ spectrum, even in the worst case of dust contamination at 150~GHz (Figure~\ref{fig:galactic_foreground}).

We also test for foreground contamination by cross-correlating the \bicep\ polarization maps with various foreground templates.
We compute the cross power spectra between \bicep\ and the \cite{odea_2012} dust models, FDS Model 8\footnote{For the FDS Model, the polarization maps are constructed by assuming 5\% 
fractional polarization split evenly between $Q$ and $U$.} \citep{finkbeiner_1999}, and the PSM dust and synchrotron predictions.
We find no statistically significant correlations between the \bicep\ maps and any of the foreground models. 

We search for compact ``point'' sources using external catalogs.
We use the same method as \citetalias{chiang_2010}, but with an updated list of catalogs including:
\planck\ \citep{planck_2013_xxviii}, \wmap\ \citep{bennett_2013}, and \acbar\ \citep{reichardt_2009}.
We find at most three (two) catalog source locations with excess power between $2\sigma$ and $3\sigma$ in the \bicep\ 100 (150) GHz map.
However, we find a similar number of $2\sigma$ to $3\sigma$ spots in signal-plus-noise simulation maps with no point sources.
Therefore we conclude that compact sources are not significant compared to the \bicep\ noise level.

As a generic test for foreground contamination we perform a $100-150$ GHz frequency jackknife.
The $\chi^2$ PTE for the $EE$, $BB$, and $EB$ spectra are 25.3\%, 99.9\%, and 84.2\%, respectively.
The PTE for $TE$ and $TB$ are 37.1\% and 65.9\%, respectively.
We conclude that the 100 and 150 GHz maps are consistent and that there is no evidence for foregrounds.

\section{Results and Data Products}
\label{s:results}

This section describes the main results on CMB bandpowers, likelihoods, and the tensor-to-scalar ratio, $r$.
All the data products described here are available online at {\bf \url{http://bicep.rc.fas.harvard.edu/bicep1\_3yr/}}.

\subsection{Bandpower Likelihood Functions}
\label{s:bandpower_likelihood}
For \bicep\ the effective number of degrees of freedom per $\ell$-bin is small so the non-Gaussianity of the likelihood is significant, especially in the lowest $\ell$-bins.
Therefore, we do not use a Gaussian likelihood approximation, but calculate bandpower likelihoods from our power spectra using the likelihood approximation of \cite{hamimeche_2008}.
The use of this approximation is a change from \citetalias{chiang_2010}, which used the then-standard offset-lognormal approximation \citep{bond_2000}.
In the course of this work we compared likelihoods on $r$ derived using the offset-lognormal distribution with those derived from a new direct likelihood (\S\ref{s:t2s:direct}), and we found the offset-lognormal likelihoods resulted in biases at the $r\sim0.1$ level.  
The $r$ estimates derived using offset-lognormal likelihoods also have significant variance among realizations from the same underlying measured value of $r$, compared to the other two methods.
The \cite{hamimeche_2008} approximation greatly reduced such bias and scatter in constraints on $r$ and we therefore adopt it as a better approximation to the true bandpower likelihood functions.

The likelihood approximation is:
\begin{equation}
-2\log\mathcal{L}(\mathbfcal{D}_b|\hat{\mathbfcal{D}}_b) = X_c\mathcal{M}^{-1}_{cc'}X_{c'},
\label{eq:HL_lik}
\end{equation}
where $\mathbfcal{D}_b$ are the model bandpowers and $\hat{\mathbfcal{D}}_b$ are the data.
In Equations~\ref{eq:HL_lik} and \ref{eq:M_HL}, indices $c$ and $c'$ run over all 54 combinations of $\ell$-bins and the six spectra ($TT$, $EE$, $BB$, $TE$, $EB$, $TB$).
Index $b$ runs only over  $\ell$-bins.
The expression for the log-likelihood is similar to a $\chi^2$ statistic, but calculated using $X_c$, a vector of bandpowers that have undergone a transformation to correct the shape of the likelihood.
\begin{equation}
\left(\begin{array}{l}
X_b^{TT}\\
X_b^{EE}\\
X_b^{BB}\\
X_b^{TE}\\
X_b^{EB}\\
X_b^{TB}\\
\end{array}\right)
 = \textrm{vecp}\left((\mathbfcal{D}^f_{b})^{1/2}\textbf{U}_bg(\textbf{D}_b)\textbf{U}_b^\dagger(\mathbfcal{D}^f_{b})^{1/2}\right),
\end{equation}
where $\mathbfcal{D}^f_{b}$ are fiducial bandpowers from the mean of \lcdm\ signal-plus-noise simulations.
The role of the fiducial model is to incorporate the bandpower covariance; $\mathbfcal{D}^f_{b}$ and $\mathcal{M}_{cc'}$ are calculated from the same simulations.
The function vecp gives the vector of unique elements in a symmetric matrix, and $\mathbfcal{D}_b, \hat{\mathbfcal{D}}_b$, and $\mathbfcal{D}^f_{b}$ are symmetric matrices constructed from the bandpowers at each $\ell$-bin, $b$. 
For example,
\begin{equation}
\mathbfcal{D}_b = \left(\begin{array}{lll}
\mathcal{D}_b^{TT} & \mathcal{D}_b^{TE} & \mathcal{D}_b^{TB} \\
\mathcal{D}_b^{TE} & \mathcal{D}_b^{EE} & \mathcal{D}_b^{EB} \\
\mathcal{D}_b^{TB} & \mathcal{D}_b^{EB} & \mathcal{D}_b^{BB} \\
\end{array}\right).
\end{equation}
The bandpowers used for this approximation are not debiased for noise or $E{\rightarrow}B$ leakage. 
Matrices $\textbf{U}_b$ and $\textbf{D}_b$ are the eigenvectors and eigenvalues of the matrix product $\mathbfcal{D}_b^{-1/2}\hat{\mathbfcal{D}}_b\mathbfcal{D}_b^{-1/2}$.
The function
\begin{equation}
g(x) = \textrm{sign}(x-1)\sqrt{2(x - \ln x - 1)}
\end{equation}
is applied to the diagonal matrix $\textbf{D}_b$ to form $g(\textbf{D}_b)$.

The bandpower covariance matrix used in Equation~\ref{eq:HL_lik} is related to the bandpower covariance matrix calculated from signal-plus-noise simulations (see \S\ref{s:ps_results}), $M_{cc'}$, by
\begin{equation}
\label{eq:M_HL}
\mathcal{M}_{cc'} = M_{cc'} + G_cG_{c'}\hat{\mathcal{D}}_c\hat{\mathcal{D}}_{c'} + S_cS_{c'}\hat{\mathcal{D}}_c\hat{\mathcal{D}}_{c'}. 
\end{equation}
The additional terms account for systematic uncertainty from absolute gain ($G_c$) and beam width ($S_c$) calibration.
Incorporating the systematic uncertainty in this way is an approximation to the likelihood obtained by introducing a systematic uncertainty nuisance parameter and marginalizing over it.
Detailed checks of this approximation will be in an upcoming paper on likelihood methods.
As in \citetalias{chiang_2010}, we use only the terms of $M_{cc'}$ that are two or fewer $\ell$-bins apart.

\subsection{Consistency with \lcdm}\label{s:check_lcdm}

To assess the consistency of our results with the \lcdm\ model, we use the bandpower likelihood described above to create a likelihood-based consistency test.
For each frequency-combined auto-spectrum ($TT$, $EE$, and $BB$) we calculate $\chi^2\equiv -2\ln\mathcal{L}$ for the theory spectrum used for the $E$-no-$B$ signal simulations\footnote{We have checked that using WMAP-9 cosmological parameters instead of WMAP-5 makes a negligible difference.}.
For cross-spectra, $TE$, $TB$, and $EB$, the Hamimeche and Lewis likelihood model only allows us to calculate a total $\chi^2$ including the related auto-spectra in the likelihood; we then subtract the auto-only $\chi^2$ to get the final statistic.
For example, $\chi^2_{TE} \equiv \chi^2_{TE+TT+EE} - \chi^2_{TT} - \chi^2_{EE}$.
For each spectrum $\chi^2$, we compute the probability to exceed (PTE) as the fraction of signal-plus-noise simulations having larger $\chi^2$ than the real data.
We list the $\chi^2$ and PTE for each spectrum in Figure~\ref{fig:data_vs_lcdm}; these values show no inconsistency with \lcdm.

\begin{figure}[t]
\resizebox{\columnwidth}{!}{\includegraphics{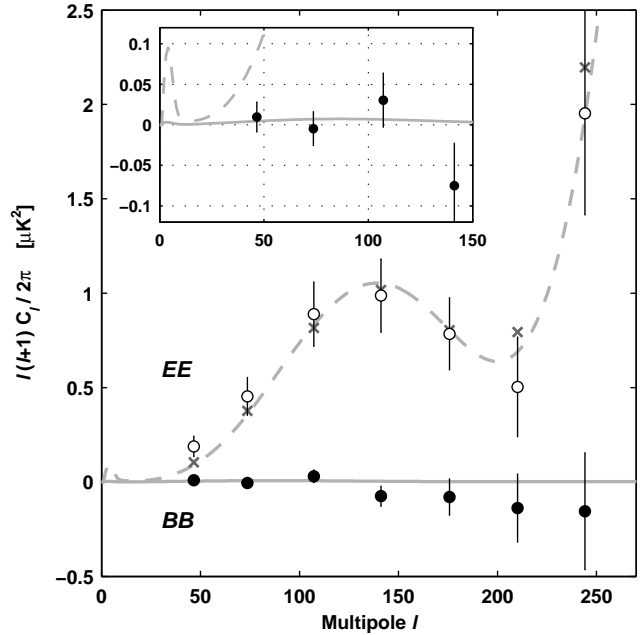}}
\caption{Close-up of the $EE$ and $BB$ spectra from Figure~\ref{fig:data_vs_lcdm}. \bicep\ measures $EE$ polarization (open circles) with high
signal-to-noise at degree angular scales (\S\ref{s:check_lcdm}). 
The $BB$ spectrum (black points) is consistent with zero. 
Theoretical $BB$ and $EE$ spectra with $r = 0.1$ are shown in solid and dashed gray lines respectively.
The gray crosses are the bandpower expectation value for the $EE$ spectrum.  
They diverge from the \lcdm\ curve because the detailed shape of the bandpower window functions. 
The inset shows the low-$\ell$ region in more detail.}
\label{fig:eebb}
\end{figure}

We also use this likelihood approximation to calculate the significance of our detection of \emode\ power (Figure~\ref{fig:eebb}).
Using only the $\ell$-bins around  the first peak of the $EE$ spectrum ($56 \leq\ell\leq 195$), we calculate the $\chi^2$ for a model with zero power to be 241. 
This corresponds to a 15$\sigma$ detection of power in the region of the first peak.
Our $EE$ detection significance using all nine $\ell$-bins is $18\sigma$.
Our $TE$ detection significance is $14\sigma$ for all $\ell$ and $8\sigma$ for the  $56\leq\ell\leq160$ region indicating the detection of superhorizon adiabatic fluctuations \citep{peiris_2003} first detected by \wmap\ \citep{kogut_2003}.

The 95\% confidence upper limits on the $BB$ spectrum (Figure~\ref{fig:pscomp}) come from applying this likelihood approximation to each $BB$ bandpower individually and excluding all other bandpowers from the calculation.
We then apply a uniform positive prior on the bandpower and integrate the resulting posterior probability distribution function (PDF) to find the limit containing 95\% of the probability.

\subsection{Constraints on Tensor-to-Scalar Ratio, $r$}
\label{s:t2s}

The primary motivation for the \bicep\ measurement of the $BB$ spectrum is to constrain the tensor-to-scalar ratio, $r$.
Following standard practice, we define $r$ as the ratio of power in primordial gravitational waves to curvature perturbations at a pivot scale $k_0=0.002$ Mpc$^{-1}$.
To model the $BB$ spectrum at a specific value, $r_*$, we simply scale the $r=0.1$ model spectrum, described in \S\ref{s:ssim}, by $(r_*/0.1)$.
This method, which uses a fixed template shape for the $BB$ spectrum and scales the amplitude, does not technically satisfy the slow roll consistency relation, $n_T = - r / 8$ \citep{kinney_1998}, but it provides a convenient and model-independent measure of sensitivity to a tensor-type $BB$ signal and is consistent with the treatments in \citetalias{chiang_2010}\ and publications from the \quiet\ Collaboration.

\begin{figure*}
\begin{center}
\resizebox{\textwidth}{!}{\includegraphics{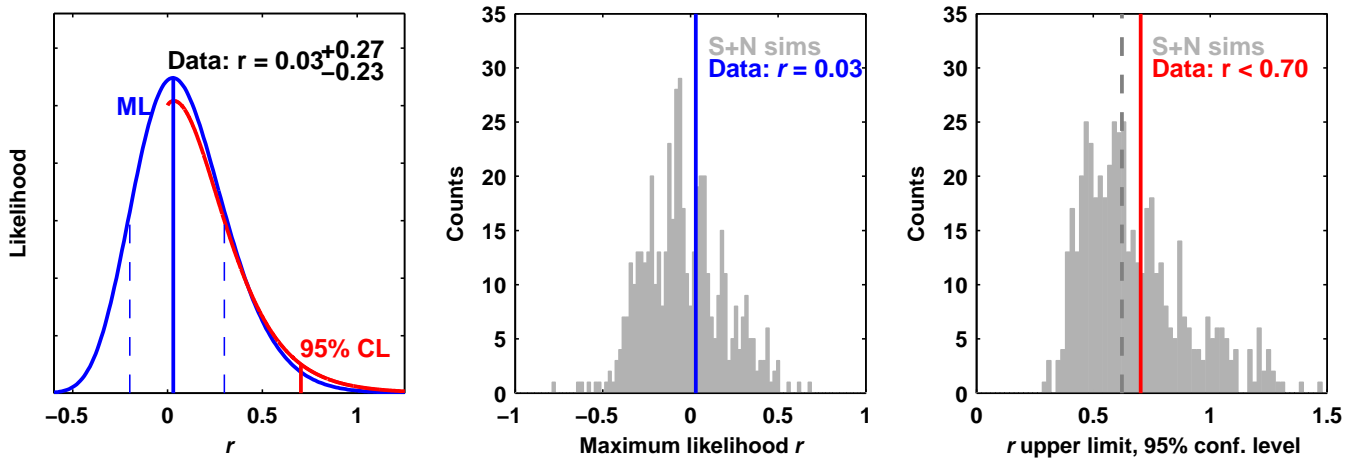}}
\caption{The likelihood for $r$ calculated from the \bicep\ $BB$ spectrum is shown in the left panel. 
The red curve comes from a direct likelihood calculation described in \S\ref{s:t2s:direct}.
The blue curve comes from an alternate calculation based on the bandpower likelihood approximation (\S\ref{s:t2s:hl}).
The maximum likelihood value and $1\sigma$ interval, $r = 0.03_{-0.23}^{+0.27}$, are shown as the blue solid and dashed lines.
A histogram of maximum likelihood $r$ values
derived from 499 signal-plus-noise simulations (with $r=0$ input) is shown in the central panel.
In the right panel, we derive 95\% confidence upper limits on $r$ from simulated likelihoods (gray histogram) and real data likelihood.  
\bicep\ obtains an upper limit of $r<0.70$ (red line), which lies within the simulated distribution. 
The gray dashed line shows the median ($r<0.65$) of the upper limits derived from simulations.}
\label{fig:r}
\end{center}
\end{figure*}

\subsubsection{Direct Likelihood Calculation}
\label{s:t2s:direct}

End-to-end signal simulations containing a tensor-type $BB$ spectrum allow us to directly compute the one-dimensional likelihood for $r$, without bandpower likelihood approximations as an intermediate step.
This method involves the definition of a quadratic estimator, 
\begin{equation}
  \rho = \alpha + \beta \sum_b \mathcal{W}_b \mathcal{D}_b^{BB},
  \label{eq:rho_def}
\end{equation}
where $\mathcal{D}_b^{BB}$ are frequency-combined $BB$ bandpowers, $\mathcal{W}_b$ are weights selected to target the \bmode\ signature of inflation, and $\alpha$ and $\beta$ are calibrated from simulations so that $\rho$ is an unbiased estimator of $r$.

The weights used to combine $BB$ bandpowers are calculated as
\begin{equation}
  \mathcal{W}_b = \sum_{b'} \mathcal{M}_{bb'}^{-1} \mathcal{A}_{b'},
  \label{eq:rho_weight}
\end{equation}
where $\mathcal{M}_{bb'}$ is the $9\times9$ $BB$ block of the bandpower covariance matrix and $\mathcal{A}_{b'}$ are signal expectation values calculated by applying the $BB$ bandpower window functions to the template $BB$ spectrum.
Note that the covariance matrix used here is calculated from signal-plus-noise simulations with a standard \lcdm\ theory spectrum and $r=0$, meaning that the estimator $\rho$ is optimized for the case where $r$ is not detected.
\bicep\ is designed to target the peak in the $BB$ spectrum occurring at $\ell \sim 80$.
As expected, the weights chosen by this method strongly emphasize the first three $\ell$-bins, which contribute 37\%, 44\%, and 15\%, respectively, to the total of $\mathcal{W}_b$. 
This choice does not bias our estimate of $r$, but merely imposes a negligible penalty to the noise of our estimator in the case of nonzero true $r$.

Next, we generate simulated maps for a range of $r$ values by combining maps from the standard $E$-no-$B$ signal-plus-noise simulations with $B$-no-$E$ signal-only $Q$ and $U$ maps that have been scaled by $\sqrt{r/0.1}$.
These simulations are inherently restricted to the physically meaningful range, $r \ge 0$.
It is necessary to add {\bmode}s to the maps, rather than simply adding a scaled $BB$ power spectrum because, while the {\bmode}s from the $B$-no-$E$ signal simulations have no correlation with the signal-plus-noise maps, the cross terms do contribute additional variance to the bandpowers.

$BB$ bandpowers are calculated from the maps, combined across frequencies, and then further summed according to the weights derived above to obtain a ``raw'' version of the $\rho$-statistic, unscaled by $\beta$.
Because the signal and noise are uncorrelated, the ensemble average of $\rho$ is linearly proportional to $r$, allowing us to fit for $\alpha$ and $\beta$.
The $BB$ bandpowers have already been debiased for contributions from instrument noise and $E \rightarrow B$ leakage, so the fit value of $\alpha$ is  small.
However, our simulations do not include the \bmode\ signal generated by gravitational lensing of {\emode}s; we correct for this by debiasing the $\rho$-statistic calculated from real data by an amount corresponding to $r=0.03$, which is the value obtained by applying the $\rho$ estimator to the expected lensing $BB$ spectrum.
Applying the calibrated $\rho$ estimator to the real \bicep\ data, we obtain $\hat{\rho} = 0.038 \pm 0.233$.
The $1\sigma$ error bar on this estimator is given by the square root of the variance of $\rho$ values simulated for the fiducial ($r=0$) model.
It is important to note that $\hat{\rho}$ is not a maximum likelihood estimate of $r$ (maximum likelihood estimates are presented in \S\ref{s:t2s:rlimit}).
Rather, it is similar to the bandpowers and error bars shown in Figure~\ref{fig:data_vs_lcdm}, which are direct measurements of power in the map, but scaled and with error estimates from simulations.

By running the \lcdm$+r$ simulations described above, we can determine the probability density of our estimator $\rho$ as a function of the input $r$.
We model this probability density function as a scaled and shifted $\chi^2$ distribution, which fits the simulated histograms well. 
The shift in the distribution can be calculated from the known noise and $E \rightarrow B$ leakage biases, which had previously been subtracted from the bandpowers; the scaling and degrees-of-freedom parameters are estimated from the mean and variance of the simulated $\rho$ values.
Including absolute gain and beam width calibration uncertainties modifies the distribution, slightly increasing its variance.

The \bicep\ likelihood function for $r$ is obtained directly by calculating the probability of obtaining the observed value, $\hat{\rho}$, as a function of model parameter $r$.
This likelihood function is shown as the red curve in the left panel of Figure~\ref{fig:r}.
The tabulated likelihood computed by this method is available as part of the \bicep\ three-year data release.
We consider it to be the most reliable description of our constraint on $r$ as it avoids bandpower likelihood approximations.
 
\subsubsection{Alternate Likelihood Calculation}
\label{s:t2s:hl}

In addition to the direct likelihood computed above, it is useful to derive an estimator that is distributed symmetrically about the true value of $r$.  To that end we  construct an alternate likelihood for $r$ based upon the bandpower likelihood approximation of \S\ref{s:bandpower_likelihood}.
This alternative makes more assumptions than the direct method but has the advantage of being defined in the unphysical region of negative $r$.
Therefore, we use it to calculate the maximum likelihood $r$ and associated 68\% confidence interval, which we allow to extend into the negative $r$ region.
We calculate this likelihood using a theory spectrum template calculated from an $r =0.1$ model, and using information from the $BB$ spectrum only.
We include the effect of gravitational lensing {\bmode}s by adding a constant lensing spectrum consistent with \lcdm\ to the theory model at every $r$.
We include the systematic uncertainty as described in \S\ref{s:bandpower_likelihood}.
The resulting maximum likelihood and minimum width 68\%  interval (uniform prior) are $r=0.03^{+0.27}_{-0.23}$ (Figure~\ref{fig:r}).

\subsubsection{Upper Limit and Confidence Intervals}
\label{s:t2s:rlimit}

\begin{figure}[b]
  \resizebox{\columnwidth}{!}{\includegraphics{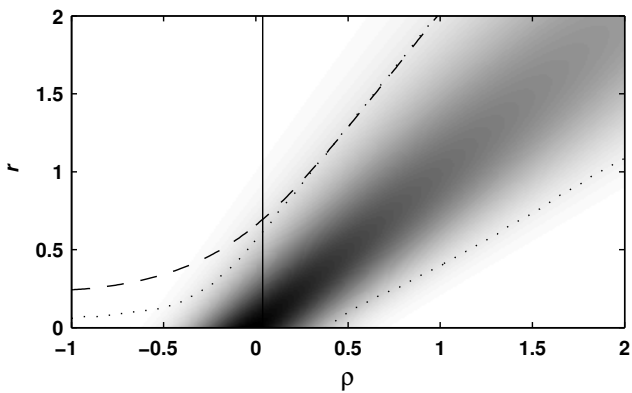}}
  \caption{Bayesian 95\% upper limit (dashed line) and Feldman-Cousins 95\% confidence interval (dotted lines) on $r$ for the \bicep\ three-year result, as a function of the value of $\rho$.  The shaded image shows the probability density of $\rho$ as a function of $r$, derived from simulations; each horizontal slice of the image yields a normalized PDF for $\rho$ given a particular theory. Vertical slices correspond to likelihood functions for $r$. The solid vertical line indicates the value of $\rho$ measured by \bicep.}
  \label{fig:prob_distribution}
\end{figure}

The left panel of Figure~\ref{fig:r} shows the \bicep\ likelihood function for $r$ calculated using both the direct method (simulation based) and alternate method (via Hamimeche \& Lewis bandpower likelihood approximation).
The most notable difference is that the alternate likelihood calculation extends to negative non-physical values of $r$.
Because of this feature, we choose to derive the maximum likelihood estimate of $r$ from the alternate likelihood calculation, as the likelihood peak will always exist, regardless of whether the data contain a high or low noise fluctuation.
For the specific case of the \bicep\ three-year results, the likelihood peaks at a slightly positive value, $r=0.03$, and the two likelihood calculations agree on the peak position; this agreement is generally quite good for all simulated results with maximum likelihood $r$ above zero.
The $1\sigma$ error bar quoted on the maximum likelihood estimate is a minimum width 68\% interval, calculated assuming a uniform prior on $r$ (positive and negative).
The center panel of Figure~\ref{fig:r} shows the distribution of maximum likelihood estimates obtained from a set of 499 simulations with input $r = 0$.

To set a 95\% confidence upper limit, we adopt a uniform prior for $r \ge 0$ only and calculate the one-sided 95\% credible interval.
This construction, which was previously used in \citetalias{chiang_2010}, as well as many other experiments in the literature, has the welcome property that it will not yield arbitrarily low (or even negative) upper limits even in the case of unlikely downward fluctuations in the data.
Since the calculation involves only the parts of the likelihood with $r \ge 0$, we can use the direct likelihood calculation, which diverges from the alternate likelihood specifically in the tails of the distribution.
The upper limit from the direct likelihood is more conservative than the same limit calculated from the alternate likelihood, both for the specific case of the \bicep\ three-year data and also for simulations of that data. We believe that the direct likelihood is more accurate, though the agreement in the region of the likelihood peak shows that the Hamimeche \& Lewis (HL) bandpower likelihoods are an excellent choice for most purposes.
The 95\% confidence upper limit from three years of \bicep\ observations is $r < 0.70$.
The right panel of Figure~\ref{fig:r} shows this limit along with the distribution of upper limits obtained from simulations. The median upper limit, a useful benchmark of experimental sensitivity, is $r < 0.65$ at 95\% confidence.
For both likelihood methods (direct and HL bandpower likelihood), we estimate the Monte-Carlo uncertainty by repeating the above calculations for the first $250$ and the last $249$ realizations separatly and find that the 95\% confidence upper limits and the maximum  likelihood estimate differ by  $\Delta r<0.04$ between each half.

The direct likelihood procedure, involving $\rho$ values calculated across a range of input $r$ models, lends itself naturally to the construction of frequentist confidence intervals.
As an alternative to the one-sided 95\% credible interval that we use for the headline upper limit on $r$, we also offer a frequentist 95\% confidence interval following the construction described in \citet{feldman_1998}.
This interval construction is chosen because it handles the physical constraint, $r \ge 0$, in a natural way.
Figure~\ref{fig:prob_distribution} shows the probability distribution for $\rho$ as a function of the theory, with both the Bayesian upper limit and the Feldman-Cousins confidence interval shown.
It is a feature of the Feldman-Cousins construction that the 95\% confidence interval for a bounded theory can be either one-sided or two-sided depending on the data; for the \bicep\ three-year result, we obtain a one-sided 95\% confidence interval, $r < 0.62$.

\section{Conclusions} \label{s:conclusions}

In summary, we present improved measurements of the degree-scale CMB polarization from \bicep. 
Compared to the previous data release \citepalias{chiang_2010}, we include 52\% more data with a corresponding decrease in statistical uncertainty.
We dramatically reduce systematic uncertainty by developing and implementing the relative gain deprojection technique.
We also implement two new likelihood calculations: a bandpower likelihood based on the existing \citet{hamimeche_2008} approximation
and a new direct simulation-based likelihood for $r$.
Both likelihoods are available as part of our data release.
We propose these methods as standards for future inflationary \bmode\ search experiments.

We support the new results with an extensive suite of consistency tests.
First, we show the new results are consistent with \citetalias{chiang_2010}.
The differences due to the change of analysis pipeline, noise model, bandpower window function calculation, and data selection are within expectation.
Second, jackknife null tests confirm the internal consistency of the data and analysis.

The most important results of \bicep\ are the CMB power spectra bandpowers.
Overall, the spectra are consistent with the \lcdm\ cosmological model.
We detect \emode\ power in the first acoustic peak at $15\sigma$, the most significant such detection to date (Figure~\ref{fig:eebb}).
We confirm the $TE$ superhorizon fluctuations, first detected by \wmap\ \citep{kogut_2003}, at $8\sigma$.
The total detection significance for non-zero $EE$ power is $18\sigma$, and $14\sigma$ for $TE$ power.

\begin{figure}[t]
  \begin{center}
    \resizebox{\columnwidth}{!}{\includegraphics{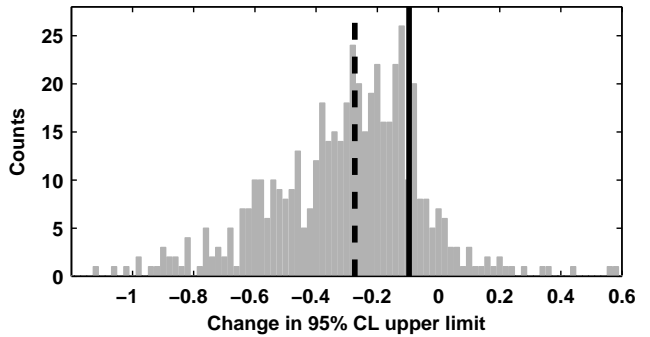}}
    \caption{Histogram of the shift in the 95\% confidence upper limit on $r$ from 
simulations upon including the additional 52\% data of the full \bicep\ observations.
A negative value indicates that the three-year upper limit is tighter than the two-year limit, but 7\% of realizations show a positive value. The dashed black line indicates the median of the distribution ($-0.27$). The solid black line indicates the value of this shift for the real data ($-0.10$).}
    \label{fig:delta_upper_limit}
  \end{center}
\end{figure}

\begin{figure*}
\begin{center}
\resizebox{\textwidth}{!}{\includegraphics{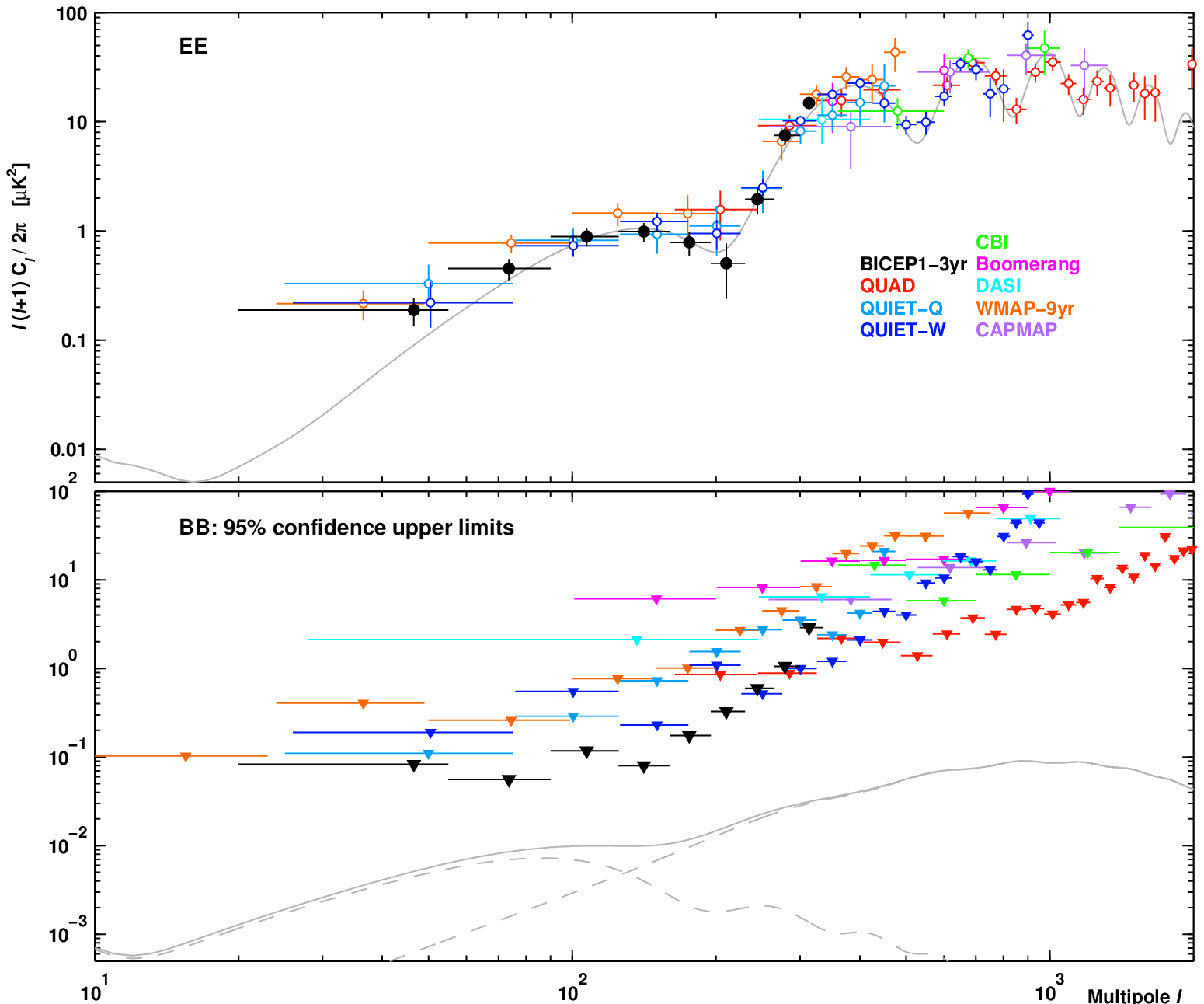}}
\caption{\bicep's  $EE$ and $BB$ power spectra complement
existing data from other CMB polarization
experiments~\citep{leitch_2005, montroy_2006,sievers_2007, bischoff_2008, brown_2009, quiet_2011, quiet_2012, bennett_2013}. For visual clarity, we only display the experiments where at least one of the $EE$ bandpowers has a center value that is
 greater than twice the distance between the center value and the
 lower end of the 68\% confidence interval. Theoretical spectra from a \lcdm\ model 
with $r = 0.1$ are shown for comparison; the $BB$ curve is the sum of 
the inflationary and gravitational lensing components.  At degree angular
scales, \bicep's constraints on $BB$ are the most powerful to date.}
\label{fig:pscomp}
\end{center}
\end{figure*}

The primary goal of \bicep\ is to search for the inflationary \bmode\ signal.
The \bmode\ spectra are consistent with zero, and we place the strongest upper limits to date in the $50 < \ell < 200$ region where the signal from inflation is expected to peak (Figure~\ref{fig:pscomp}).
We also report this result as a constraint on the tensor-to-scalar ratio, $r$.
This constraint is $r=0.03^{+0.27}_{-0.23}$ (68\% CI) or $r<0.70$ (95\% CL).
The corresponding upper limit from \citetalias{chiang_2010} is $r<0.72$.

One might naively expect the upper limit to improve by a larger factor when adding 52\% more data.
We have confirmed that the relatively small decrease is not a result
of the set of analysis refinements discussed in \S\ref{s:consistency}.
The change in mapmaking pipeline, noise model,
bandpower window function calculation, and deprojection each produce small shifts in bandpowers and error bars
as described above.  But when we apply all of these changes together to the original~\citetalias{chiang_2010}
data set, the resulting $r$ constraint derived using the same offset-lognormal likelihood approximation
as \citetalias{chiang_2010} is $r<0.71$.
In other words, the net effect of these changes on the $r$ upper limit is close to zero for this dataset.

The relatively small decrease in the new upper limit is explained by two factors.
First, the offset-lognormal likelihood approximation used in \citetalias{chiang_2010} resulted 
in a negative bias on the upper limit for those specific $BB$ bandpowers (we find that the $r$ constraint 
derived from offset-lognormal bandpower likelihoods is biased low for some cases and high for others).
Applying the more accurate direct likelihood calculation to the reanalyzed \citetalias{chiang_2010} dataset
shifts the upper limit in this case from $r<0.71$ to $r<0.80$.
Second, upon including the new data in this analysis the upper limit fluctuates
somewhat high compared to the average of simulations.
With an $r=0$ input model, simulated datasets run through our final analysis yield upper limits on $r$ that decrease
by a median of 0.27 when including the additional data of the full three years (Figure~\ref{fig:delta_upper_limit}).
The corresponding decrease seen in the real data is only 0.10 (from 0.80 to 0.70). 
Although this decrease is smaller than average it is not an unlikely result;
17\% of the simulations saw even less of a decrease, and
in 7\% of the simulations the upper limit actually increases when adding the additional data.

Interesting constraints can be placed on cosmic birefringence from the \bicep\ $TB$ and $EB$ spectra, which are predicted to by zero by the \lcdm\ model.
This topic will be explored in detail in \cite{kaufman_2013}.

Measurement of CMB \bmode\ polarization remains the most promising approach for testing the inflationary paradigm.
\bicep\ has provided the lowest upper limits on inflationary $B$-modes to date.
In this data release, we demonstrate a deprojection technique that will enable future experiments to cope with the increasingly important temperature-to-polarization leakage and develop a direct likelihood calculation for converting bandpower results into constraints on $r$ without approximations.
\biceptwo\ has completed three years of observation with an order of magnitude better mapping speed at 150~GHz than \bicep.
\keckarray\ operations are ongoing, with two full years of observation completed by November 2013 \citep{ogburn_2012,aikin_2010}.
\bicepthree\ will begin observing in 2014--2015.
Measurements of \bmode\ polarization from these and other experiments, using the new analytical tools we have demonstrated here,
have the potential to test inflationary cosmology with unprecedented precision.

\acknowledgements

\bicep\ was supported by NSF Grant No. OPP-0230438, Caltech President's
Discovery Fund, Caltech President's Fund PF-471, JPL Research and
Technology Development Fund, and the late J.~Robinson.  
This analysis was supported in part by NSF CAREER award No.  AST-1255358 and
the Harvard College Observatory, and J.M.K. acknowledges support from
an Alfred P. Sloan Research Fellowship.
B.G.K acknowledges support from NSF PECASE Award No. AST-0548262. 
We thank the
South Pole Station staff for helping make our observing seasons a
success. 
 We also thank our
colleagues in \acbar, \boom, \QUAD, \bolocam, \spt, \wmap\  and \planck\ for
advice and helpful discussions, and Kathy Deniston and Irene Coyle
for logistical and administrative support.  
We thank Patrick Shopbell for computational support at Caltech 
and the FAS Science Division Research Computing Group at Harvard University 
for providing support to run all the computations for this paper on the Odyssey cluster.

\bibliographystyle{apj}
\bibliography{2012_bicep_cmb_3yr}

\end{document}